\documentclass[twocolumn]{aastex63}

\usepackage{lipsum} 

\usepackage{graphicx}
\usepackage{tikz}
\usepackage[export]{}

\usepackage{amsmath}

\usepackage{fontenc}
\usepackage{cancel}
\usepackage{tensor}
\renewcommand\vec{\mathbf}
\newcommand*{\tensndd}[1]{\overline{\vec{#1}}}
\newcommand*{\tensrdd}[1]{\overline{\overline{\vec{#1}}}}
\newcommand*{\tensnd}[1]{\overline{\mathbb{#1}}}

\newcommand*{\uvec}[1]{\hat{\vec{#1}}}

\DeclareMathSymbol{\mh}{\mathord}{operators}{`\-}
\received{December -, 2021}
\revised{-- --, 2022}
\accepted{-- --, 2022} 
\submitjournal{APJ}

\shorttitle{Energy Transfer and Turbulent Reconnection}
\shortauthors{J.A. Agudelo Rueda et al.}

\graphicspath{{./}{figures/}}

\begin{document}

\title{Energy transport during 3D small-scale reconnection driven by anisotropic plasma turbulence}

\correspondingauthor{Jeffersson Andres Agudelo Rueda}
\email{jeffersson.agudelo.18@ucl.ac.uk}

\author[0000-0001-5045-0323]{Jeffersson A Agudelo Rueda}
\affiliation{Department of Physics and Astronomy, Dartmouth College, Hanover, NH, USA}
\affiliation{Mullard Space Science Laboratory, UCL,
Holmbury St. Mary, Dorking, Surrey, RH5 6NT, UK}

\author[0000-0002-0497-1096]{Daniel Verscharen}
\affiliation{Mullard Space Science Laboratory, UCL,
Holmbury St. Mary, Dorking, Surrey, RH5 6NT, UK}

\author[0000-0002-0622-5302]{Robert T. Wicks}
\affiliation{Department of Mathematics, Physics and Electrical Engineering,  Northumbria University,
Newcastle upon Tyne, NE1 8ST, UK}

\author[0000-0002-5982-4667]{Christopher J. Owen}
\affiliation{Mullard Space Science Laboratory, UCL,
Holmbury St. Mary, 
Dorking, Surrey, RH5 6NT, UK}

\author[0000-0003-3623-4928]{Georgios Nicolaou}
\affiliation{Mullard Space Science Laboratory, UCL,
Holmbury St. Mary, 
Dorking, Surrey, RH5 6NT, UK}

\author[0000-0003-2672-9249]{Kai Germaschewski}
\affiliation{Space Science Center,
University of New Hampshire,
Durham NH 03824, USA}

\author[0000-0003-3623-4928]{Andrew P. Walsh}
\affiliation{European Space Astronomy Centre, Urb. Villafranca del Castillo, E-28692 Villanueva de la Cañada, Madrid, Spain}

\author[0000-0002-1682-1212]{Ioannis Zouganelis}
\affiliation{European Space Astronomy Centre, Urb. Villafranca del Castillo, E-28692 Villanueva de la Cañada, Madrid, Spain}

\author[0000-0002-5999-4842]{Santiago Vargas Dom\'inguez}
\affiliation{Universidad Nacional de Colombia, Observatorio Astronómico Nacional, Ed. 413 Bogotá, Colombia}
\begin{abstract}

Energy dissipation in collisionless plasmas is a longstanding fundamental physics problem. Although it is well known that magnetic reconnection and turbulence are coupled and transport energy from system-size scales to sub-proton scales, the details of the energy distribution and energy dissipation channels remain poorly understood. Especially, the energy transfer and transport associated with three dimensional (3D) small-scale reconnection that occurs as a consequence of a turbulent cascade is unknown. We use an explicit fully kinetic particle-in-cell code to simulate 3D small scale magnetic reconnection events forming in anisotropic and Alfvénic decaying turbulence. We identify a highly dynamic and asymmetric reconnection event that involves two reconnecting flux ropes. We use a two-fluid approach based on the Boltzmann equation to study the spatial energy transfer associated with the reconnection event and compare the power density terms in the two-fluid energy equations with standard energy-based damping, heating and dissipation proxies. Our findings suggest that the electron bulk flow transports thermal energy density more efficiently than kinetic energy density. Moreover, in our turbulent reconnection event, the energy-density transfer is dominated by plasma compression. This is consistent with turbulent current sheets and turbulent reconnection events, but not with laminar reconnection. 
\end{abstract}


\section{Introduction} 
\label{sec:intro}

The solar wind in the inner heliosphere is a weakly collisional turbulent plasma in which the energy is transported from large ($\sim 10^9 $ km) to small ($\sim 10^{-1} $ km) scales via an active turbulent cascade \citep{coleman1968turbulence, marsch1990spectral}. Although the collisionless nature of the solar wind precludes classical viscous dissipation of these turbulent fluctuations, the non-adiabatic evolution of the solar wind \citep{gazis1982voyager,matteini2007evolution,hellinger2011heating} suggests the action of local heating mechanisms \citep{barnes1968collisionless,goldstein2015kinetic}. The plasma-physics processes responsible for this heating are not fully understood yet. The observed velocity distribution function of the solar wind species often exhibits non-thermal features e.g. \citep{marsch1982solar,feldman1975solar,feldman1978characteristic,mccomas1992solar}. 

Important progress has been made to understand heating and energy dissipation e.g. \citep{gary1999collisionless,howes2017diagnosing,klein2017diagnosing,matthaeus2020pathways}. Landau damping, ion cyclotron damping and stochastic heating are considered collisionless dissipation mechanisms that transfer energy from the electromagnetic field to the plasma particles \citep{marsch2003ion, kasper2008hot, chandran2010perpendicular,chandran2013stochastic}. The dissipation occurs predominantly in intermittent structures which form in plasma turbulence \citep{matthaeus1999turbulence,kiyani2015dissipation}. 

Like turbulence, magnetic reconnection is a process that emerges on a broad range of scales and under a large variety of plasma conditions. Magnetic reconnection occurs when magnetic structures form regions in which the frozen-in condition is locally broken allowing the exchange of particles between the magnetic structures \citep{hesse1988theoretical,schindler1988general}. 

Magnetic reconnection and turbulence are closely linked. Magnetic reconnection self-consistently occurs as a consequence of the turbulent cascade \citep{servidio2010statistics, loureiro2020nonlinear,agudelo2021three} and turbulence emerges in current sheets, exhaust flows, electron streamers, and shocks associated with reconnection events \citep{pucci2017properties,kowal2017statistics,lapenta2020local}. During magnetic reconnection, plasma particles are heated and accelerated while the magnetic field topology changes \citep{pontin2011three,zweibel2016,lazarian20203d}. 

The role of magnetic reconnection for the evolution of energy in collisionless plasmas is unclear. Although magnetic reconnection transports energy from large to small scales \citep{sundkvist2007dissipation,franci2017magnetic, loureiro2020nonlinear}, the details of the energy transport across scales and the role of reconnection in the turbulent cascade are a matter of ongoing research \citep{loureiro2017role,franci2017magnetic,PhysRevE.104.065206}. The energy transfer between fields and particles as well as the transfer between kinetic and thermal degrees of freedom during reconnection are the key objectives of this research area.

The energy transfer and transport associated with magnetic reconnection has been addressed by previous studies that focus on idealized 2D Harris current-sheet reconnection \citep{yin2001hybrid,schmitz2006kinetic,wang2015comparison,pezzi2021dissipation}, 3D laminar collisionless reconnection in the context of magnetospheres \citep{wang2018electron} and 2D reconnection in turbulent plasma \citep{fadanelli2021energy}. In this work, we use particle-in-cell (PIC) simulations to study the energy transfer associated with 3D small-scale magnetic reconnection that self-consistently occurs as a consequence of an anisotropic turbulent cascade. In section \ref{sec:energy_dist}, we present our theoretical framework to study the energy transfer and transport in our plasma simulations. In section \ref{sec:Initialization_sim} we present our simulation results emphasizing the presence of agyrotropy in section \ref{sec:Agirotropy} and the energy distribution in section \ref{sec:energy_distri}. In section \ref{sec:discussion}, we discuss the implications of our results and in section \ref{sec:conclusions} we provide conclusions. 


\section{Energy transfer and transport} 
\label{sec:energy_dist}

The total energy in a closed volume of plasma is partitioned amongst the particles and the electromagnetic fields. The bulk kinetic energy density of the particle species $s$ is associated with the first velocity moment of the particle velocity distribution function $f_{s} = f_{s}(\vec{x},\vec{v},t)$ and therefore with the bulk flux of the particles. The thermal energy density is associated with the second velocity moment and thus the pressure of the particles. The evolution of $f_{s}$ follows the Boltzmann equation 

\begin{equation}
    \frac{\partial f_{s}}{\partial t} + \vec{v} \cdot \nabla f_{s} + \frac{q_{s}}{m_{s}} \left(\vec{E} + \vec{v} \times \vec{B} \right) \cdot \nabla_{v}f_{s}= \left( \frac{\partial f_{s}}{\partial t} \right)_{c},
    \label{eqn:boltz}
\end{equation}

\noindent where $\vec{v}$ is the velocity, $\vec{E}$ is the electric field, $\vec{B}$ is the magnetic field, $q_{s}$ is the charge and $m_{s}$ is the mass of a particles. The term $(\partial f_{s} / \partial t )_{c}$ on the right-hand side represents the change in the distribution function due to collisions. This term includes individual correlations between fields and particles, based on the particles' individual Coulomb potentials \citep{klimontovich1997physics}. To study the energy transport we derive a set of energy equations based on the Boltzmann equation (\ref{eqn:boltz}). We first define the density

\begin{equation}
n_{s}\equiv \int f_{s} d^{3}v,
\label{eqn:density}
\end{equation}

\noindent the bulk velocity 
\begin{equation}
\vec{u}_{s} \equiv  \frac{1}{n_{s}} \int f_{s}\vec{v} d^{3}v,
\label{eqn:bulkvelocity}
\end{equation}

\noindent and the pressure tensor 

\begin{equation}
\tensndd{P}_{s} \equiv m_{s}\int f_{s}(\vec{v}-\vec{u}_s)(\vec{v}-\vec{u}_{s}) d^{3}v,
\label{eqn:pressure_tensor}
\end{equation}

\noindent where $(\vec{v}-\vec{u}_s)(\vec{v}-\vec{u}_{s})$ is the dyadic product. We define the heat flux vector 

\begin{equation}
\vec{h}_{s} \equiv \frac{1}{2} m_{s}\int f_{s}(\vec{v}-\vec{u}_{s})\cdot(\vec{v}-\vec{u}_s)(\vec{v}-\vec{u}_{s}) d^{3}v.
\label{eqn:heatvector}
\end{equation}

\noindent We define the first moment of the collision term in Eq. (\ref{eqn:boltz}) as

\begin{equation}
\vec{\Xi}^{1} = \int \vec{v}\left( \frac{\partial f_{s}}{\partial t} \right)_{c} d^{3}v,
\label{eqn:collint}
\end{equation}

\noindent and the second moment as 

\begin{equation}
\tensndd{\Xi}^{2} = \int \vec{v}\vec{v}\left( \frac{\partial f_{s}}{\partial t} \right)_{c} d^{3}v,
\label{eqn:collint2}
\end{equation}

\noindent With these definitions, we compute the first and second moments of Eq. (\ref{eqn:boltz}) (see Appendix \ref{app:Energy_equa} for details). The first moment of Eq. (1) yields the kinetic energy equation

\begin{equation}
    \frac{d \varepsilon_{s}^{k}}{d t} + \vec{u}_{s} \cdot \nabla \cdot \tensndd{P}_{s} + \varepsilon_{s}^{k}\nabla\cdot\vec{u}_{s} - q_{s}n_{s}(\vec{u}_{s} \cdot \vec{E}) = \Xi^{k}_{s},
    \label{eqn:firstmomener_text}
\end{equation}

\noindent where $d/dt=\partial /\partial t + (\vec{u}_{s} \cdot \nabla)$ is the total time derivative, 

\begin{equation}
\varepsilon^{k}_{s} = \frac{1}{2} n_{s}m_{s}(\vec{u}_{s} \cdot \vec{u}_{s}),
\end{equation}

\noindent is the kinetic energy density and

\begin{equation}
\Xi^{k}_{s} =  m_{s} \vec{u}_{s} \cdot \vec{\Xi}_{s}^{1}
\end{equation}

\noindent represents the irreversible kinetic energy transfer. The terms $\vec{u}_{s} \cdot \nabla \cdot \tensndd{P}_{s}$, $\varepsilon_{s}^{k}\nabla\cdot\vec{u}_{s}$ and the advective term $(\vec{u}_{s}\cdot \nabla)\varepsilon^{k}_{s}$ are associated with the term $\vec{v} \cdot \nabla f_{s}$ in Eq. (\ref{eqn:boltz}). Therefore, these terms represent kinetic energy-density transport due to the free streaming of particles. Conversely, the term $-q_{s}n_{s}(\vec{u}_{s} \cdot \vec{E})$, associated with the electric field, represents the energy-density transfer between particle bulk flows and fields.  

The second moment of Eq. (\ref{eqn:boltz}) yields the thermal energy equation

\begin{equation}
\frac{d \varepsilon_{s}^{th}}{d t} + \nabla \cdot \vec{h}_{s} + \nabla\vec{u}_{s}:\tensndd{P}_{s} + \varepsilon_{s}^{th}\nabla \cdot \vec{u}_{s} = \Xi^{th}_{s},
\label{eqn:secondmomener_text}
\end{equation}

\noindent where 

\begin{equation}
\varepsilon^{th}_{s} = \frac{1}{2} Tr (\tensndd{P}_{s})
\end{equation}

\noindent is the thermal energy density and 

\begin{equation}
\Xi^{th}_{s} =  -m_{s} \vec{u}_{s} \cdot \vec{\Xi}^{1} + \frac{m_{s}}{2}Tr\left(\tensndd{\Xi}^{2}\right)
\end{equation}

\noindent represents the irreversible the thermal energy transfer. The term $Tr$ stands for the trace of the tensor and $\nabla\vec{u}_{s}:\tensndd{P}_{s}$ is the double contraction of the strain tensor $\nabla\vec{u}_{s}$ and $\tensndd{P}_{s}$. The terms $\nabla \cdot \vec{h}_{s}$, $ \nabla\vec{u}_{s}:\tensndd{P}_{s}$ and $\varepsilon_{s}^{th}\nabla \cdot \vec{u}_{s}$, associated with $\vec{v} \cdot \nabla f_{s}$ in Eq. (\ref{eqn:boltz}), represent thermal energy-density transport due to the free streaming of particles.  

The terms on the left-hand sides of Eqs. (\ref{eqn:firstmomener_text}) and (\ref{eqn:secondmomener_text}) describe collisionless processes whereas the terms on the right-hand sides describe collisional processes in the plasma which generate an increase in the plasma entropy.

{Equations (\ref{eqn:firstmomener_text}) and (\ref{eqn:secondmomener_text}) alone do not capture total energy conservation because they do not account for the rate of change in the electromagnetic energy density $\partial \varepsilon^{em}/\partial t$, nor for the electromagnetic energy flux $\nabla \cdot \vec{S}$, where} 
\begin{equation}
\varepsilon^{em} = \frac{1}{2}\left( \frac{1}{\mu_{0}} \vec{B}\cdot\vec{B} + \epsilon_{0}\vec{E}\cdot\vec{E} \right)
\end{equation}
{
is the electromagnetic energy density and $\vec{S}=\vec{E}\times\vec{B}/\mu_{0}$ is the Poynting vector. The expression that accounts for these changes is Poynting's theorem}  
\begin{equation}
\frac{\partial \varepsilon^{em}}{\partial t} + \nabla \cdot \vec{S} +\vec{J}\cdot\vec{E}=0.
\label{eqn:poynting_theo}
\end{equation}
{Nevertheless, Equations (\ref{eqn:firstmomener_text}) and (\ref{eqn:secondmomener_text}) are exact in their description of the kinetic and thermal energy-density transfer and transport, as well as the energy-density exchange between fields and particles.}

Before tackling the energy transfer problem, we explicitly define the following three terms, which are often used interchangeably in the literature:

\begin{itemize}
    \item \emph{Heating} is any increase in $\varepsilon^{th}_{s}$, and \emph{cooling} is any decrease in $\varepsilon^{th}_{s}$. Heating can be {either reversible or irreversible}.
    \item  \emph{Damping} is any decrease in $\varepsilon^{em}$, and \emph{growth} is any increase in $\varepsilon^{em}$. Damping/growth can be {either reversible or irreversible}.
    \item \emph{Dissipation} is any irreversible energy transfer {leading to an increase in} $\varepsilon^{th}_{s}$.
\end{itemize}   

Dissipation corresponds to an increase in entropy of the velocity distribution function, which is challenging to quantify directly both in space measurements and in simulations. Nonetheless, recent studies \citep{pezzi2019energy,matthaeus2020pathways,pezzi2021dissipation} show that in collisionless plasmas  \emph{energy-based dissipation proxies} such as the Zenitani parameter \citep{zenitani2011new}

\begin{equation}
D_{z,s}=\vec{J}\cdot \left( \vec{E} + \vec{u}_{s}\times \vec{B} \right) - n_{s}q_{s}(\vec{u}_{s}\cdot \vec{E}),
\label{eqn:zenitani}
\end{equation}

and the strain-pressure interaction $\nabla\vec{u}_{s}:\tensndd{P}_{s}$ \citep{yang2017energy} {are spatially correlated with dimensionless measures of non-thermal distribution functions \citep{kaufmann2009boltzmann,greco2012inhomogeneous,liang2019decomposition} and plasma agyrotropy \citep{scudder2008illuminating}}. In Eq.~(\ref{eqn:zenitani}), $\vec{J}=\sum_{s=i,e}{q_{s}n_{s}\vec{u}_{s}}$ is the electric current density.

These energy-based dissipation proxies are effectively \emph{power density} terms derived from the left-hand sides of our Eqs. (\ref{eqn:firstmomener_text}) and (\ref{eqn:secondmomener_text}). According to our definitions, $D_{z,s}$ is a damping measure since it quantifies the energy transfer from the electromagnetic fields into bulk kinetic energy and vice versa. 

The strain-tensor interaction has gyrotropic and agyrotropic contributions. We de-compose the pressure tensor as ${P}_{ij,s}={p}_{s}\delta_{ij}+{\Pi}_{ij,s}$, where 

\begin{equation}
p_{s}=\sum_{i=1}^{3} P_{ii,s}/3
\label{eqn:pi_pressure_d}
\end{equation}

\noindent is the isotropic scalar pressure and  

\begin{equation}
\Pi_{ij,s} = (P_{ij,s} + P_{ji,s})/2 - p_{s}\delta_{ij}
\label{eqn:PI_pressure_of}
\end{equation}
 
\noindent is the deviatoric pressure. Likewise, the strain-rate tensor $\nabla\vec{u}_{s}$ can be expressed as $\nabla{u}_{ij,s} = \theta_{s} \delta_{ij}/3 + D_{ij,s}$, where $\theta_{s}=\nabla\cdot\vec{u}_{s}$ represents the dilatation term and 

\begin{equation}
D_{ij,s}=\frac{1}{2}\left(\frac{\partial u_{i,s}}{\partial x_{j}} + \frac{\partial u_{j,s}}{\partial x_{i}}\right) - \frac{1}{3}\theta_{s}\delta_{ij}, 
\end{equation}

\noindent represents the symmetric traceless strain-rate tensor \citep{yang2017energy}. Thus, the strain-tensor interaction, which is a heating/cooling proxy according to our definitions, is

\begin{equation}
\nabla\vec{u}_{s}:\tensndd{P}_{s} = p_{s}\theta_{s} + \Pi_{ij,s}D_{ij,s},
\label{eqn:double_duP}
\end{equation}

\noindent where the first terms on the right-hand side is known as $p\theta_{s}$ and the second term on the right-hand side is known as $Pi \mh D_{s}$\citep{yang2017energy}. For comparison with previous studies \citep{pezzi2021dissipation,bandyopadhyay2020statistics}, in section \ref{sec:about_dissi}, we compute $D_{ij,s}$, $p_{s}\theta_{s}$ and $\Pi_{ij,s}D_{ij,s}$ and compare them with the energy transfer and transport terms, $-n_{s}q_{s}(\vec{u}_{s}\cdot\vec{E})$ and $\nabla\vec{u}_{s}:\tensndd{P}_{s}$, in Eqs. (\ref{eqn:firstmomener_text}) and (\ref{eqn:secondmomener_text}). 


\section{Simulation results}
\label{sec:Initialization_sim}

\subsection{Simulation setup}

We use the explicit Plasma Simulation Code \citep[PSC,][]{germaschewski2016plasma} to simulate anisotropic Alfvénic turbulence in an ion-electron plasma in the presence of a constant background magnetic field {$\vec{B}_{0} = B_{0}\uvec{z}$}. The simulation domain is an elongated box of size $L_{x} \times L_{y} \times L_{z} = 24d_{i}\times24d_{i}\times125d_{i}$ with spatial resolution $\Delta x =\Delta y = \Delta z =  0.06d_{i}$, where $d_{i}=c/\omega_{pi}$ is the ion inertial length, $c$ is the speed of light, $\omega_{pi}=\sqrt{n_{0}q_{i}^{2}/m_{i}\epsilon_{0}}$ is the ion plasma frequency and $n_{0}$ is the constant initial ion density. The background ion Alfvén speed ratio in our simulations is $v_{A,i}/c=0.1$, where $v_{A,i}=B_{0} / \sqrt{\mu_{0}n_{0}m_{i}}$ is the ion Alfvén speed. The number of macro particles per cell is $100$ ions and $100$ electrons. We use a mass ratio of $m_{i}/m_{e} = 100$ so that $d_e = 0.1 d_{i}$. We set the initial thermal-to-magnetic energy ratio $\beta_{s}=2 n_0 \mu_{0} k_{B}T_{s}/B_{0}^{2} = 1$ where $T_{s}$ is the temperature of species $s$ and $k_{B}$ is the Boltzmann constant. The details of the simulation setup and the overall simulation results are presented by \citet{agudelo2021three}, where the authors report a reconnection event that involves two reconnecting flux-ropes.



\subsection{Reconnection event overview}

\begin{figure*}
\centering
\begin{tikzpicture}
    \draw (0, 0) node[inner sep=0] {\includegraphics[width=1.0\linewidth]{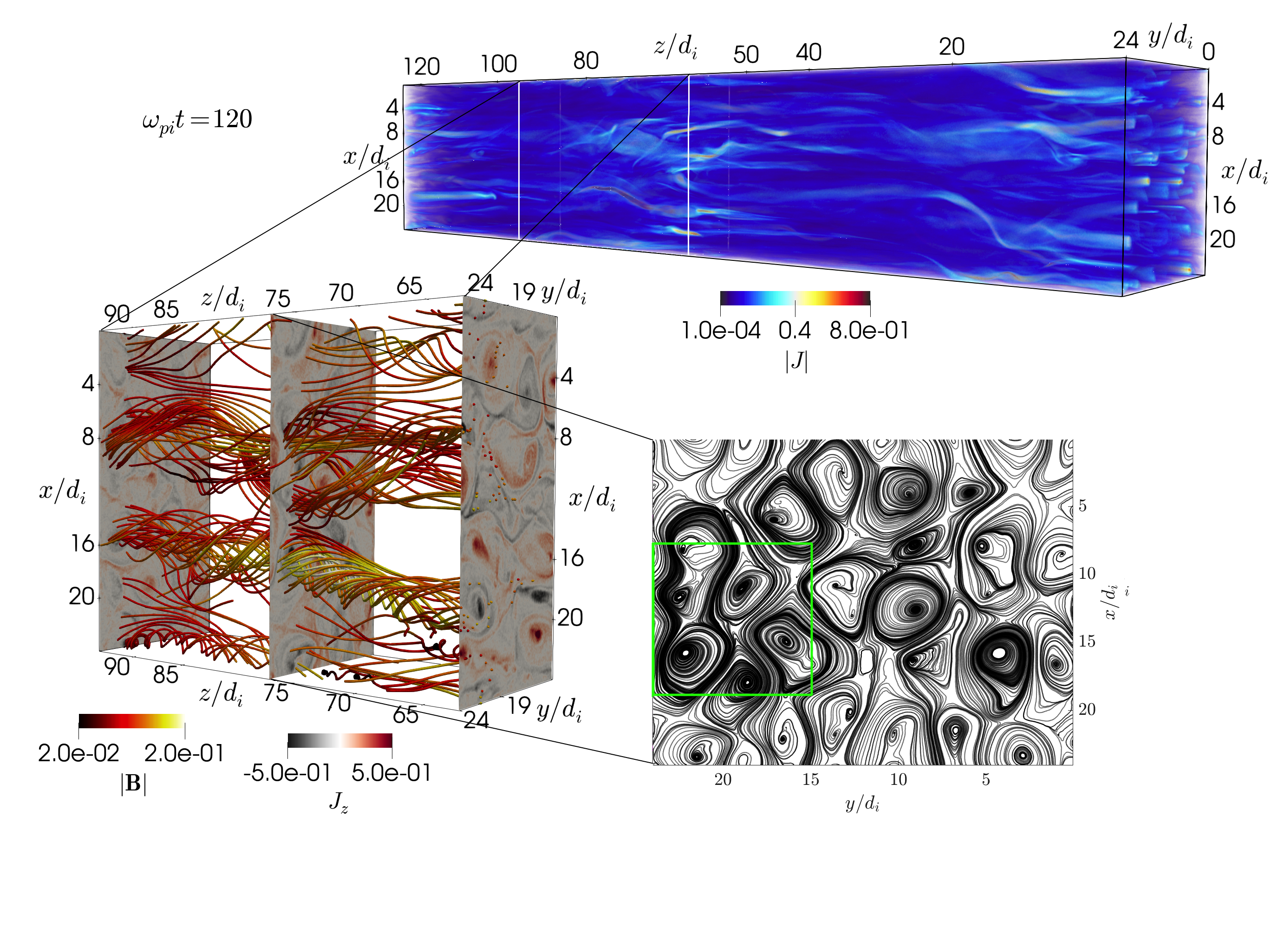}};
    \draw (-4.2, 5.7) node {a)};
    \draw (-7.3, 2.8) node {b)};
    \draw (0.5, 0.8) node {c)};
\end{tikzpicture}
\caption{Spatial context of the reconnection event within the simulation domain. Panel a) shows a volume rendering of $|\vec{J}|$. Panel b) shows the 3D magnetic field lines color-coded with $|\vec{B}|$. On the vertical cuts in panel b) we show $J_{z}$. Panel c) shows the magnetic field lines in the $xy$-plane. The black contours show in-plane magnetic-field lines. The green square highlights the size and position of the region for the energy analysis.}
\label{fig:three_composition}
\end{figure*}
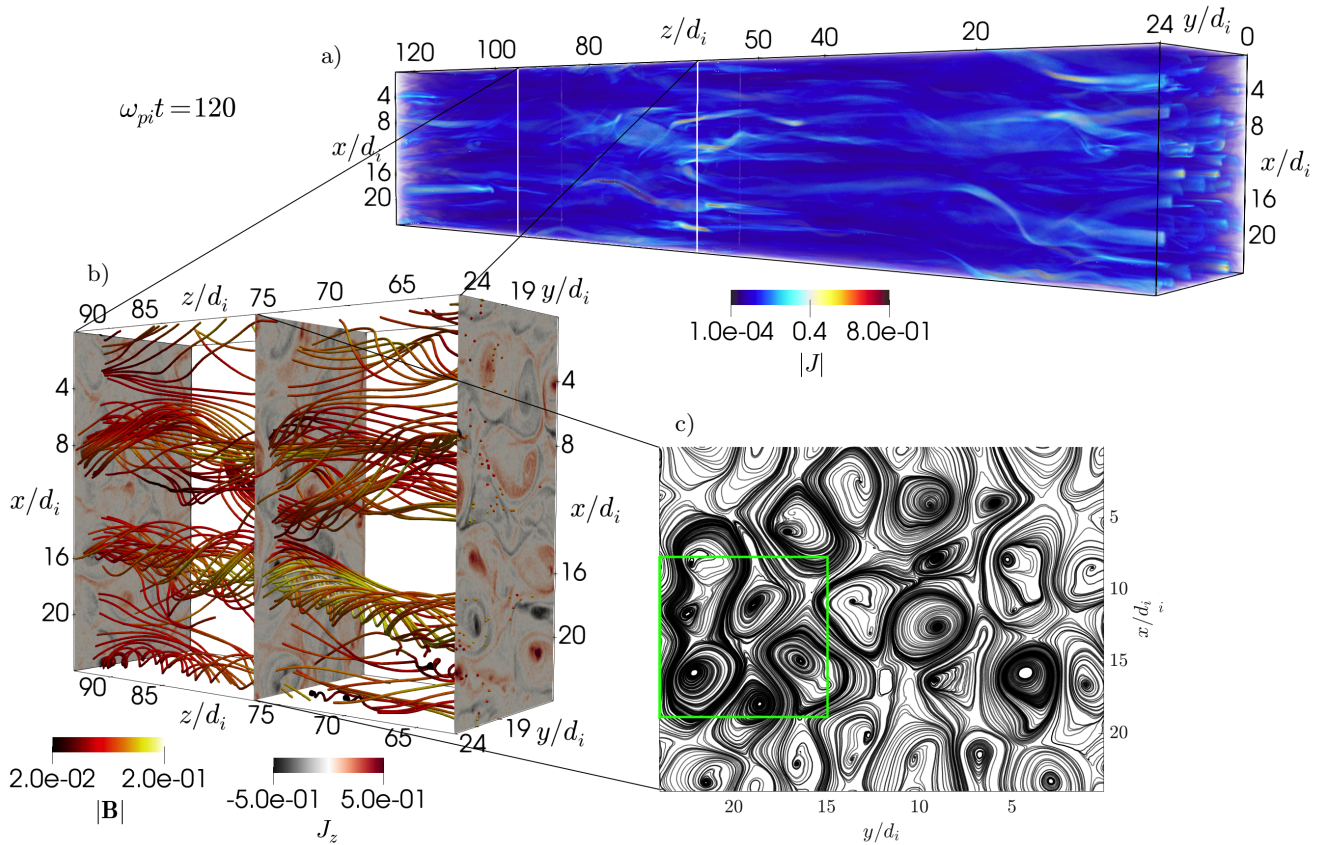

Panel a) in Figure \ref{fig:three_composition} shows the volume rendering of the current density in our simulation domain at the simulated time $t=120 \omega_{pi}^{-1}$. Current filaments that form in the turbulent cascade are mostly elongated along the direction of the background magnetic field. At this time in the simulation, we apply the set of indicators presented by \citet{agudelo2021three} to identify and locate reconnection sites. We select one reconnection event that involves two reconnecting flux-ropes as shown in panel b) of Figure \ref{fig:three_composition} where the magnetic-field lines are color-coded with $|\vec{B}|$. The magnetic flux-ropes contain an intense magnetic field, especially the lower flux-rope which is more twisted and has a smaller radius than the upper flux-rope. Conversely, the magnetic field between the flux-ropes is weak. The cuts in panel b) show $J_{z}$ in the $xy$ simulation plane. For our analysis of this event, we apply a 2D cut in the $xy$-plane at $z=77 d_{i}$. Panel c) of Figure \ref{fig:three_composition} shows the magnetic-field lines of the field components in the $xy$-plane, i.e., $(B_{x},B_{y})$ as black contours. Panel c) illustrates the complexity of the magnetic topology in the region of interest. For our energy analysis we select a volumetric sub-region of size $10$ $d_{i}^{3}$ centered around the identified reconnecting region. The green squared in panel c) highlights the intersection of the selected sub-region with the central 2D cut from panel b).

Even though the background field is in the z-direction, the current structures are not exactly aligned with the z-direction. Instead, the geometric features of the reconnection event are aligned with the plane perpendicular to the current sheet that sustains the magnetic gradient. Therefore, we determine a reference frame that is aligned with the main axis of the current sheet. We determine the direction of the main axis of the current-sheet by 3D rendering $J_{z}$ and measuring the inclination of the coherent structure that crosses the point $x=13.5 d_{i}$ and  $y=21.5 d_{i}$ in the $xy$-plane. We then apply a coordinate transformation from the reference frame (RF) ($x,y,z$) to a new RF ($r,p,a$) aligned with the main axis of the current-sheet. The unit vectors of this RF are ($\uvec{r},\uvec{p},\uvec{a}$). In this RF $\uvec{a}$ is anti-parallel to the main axis of the current-sheet, $\uvec{p}$ is an arbitrary vector in the plane perpendicular to $\uvec{a}$, and $\uvec{r}$ is the vector that completes the right-handed coordinate system. Since the components $r$ and $p$ are in the plane perpendicular to the current structure, we denote them as the in-plane components. 

\begin{figure*}
\centering
\begin{tikzpicture}
    \draw (0, 0) node[inner sep=0] 
    {\includegraphics[width=1\linewidth]{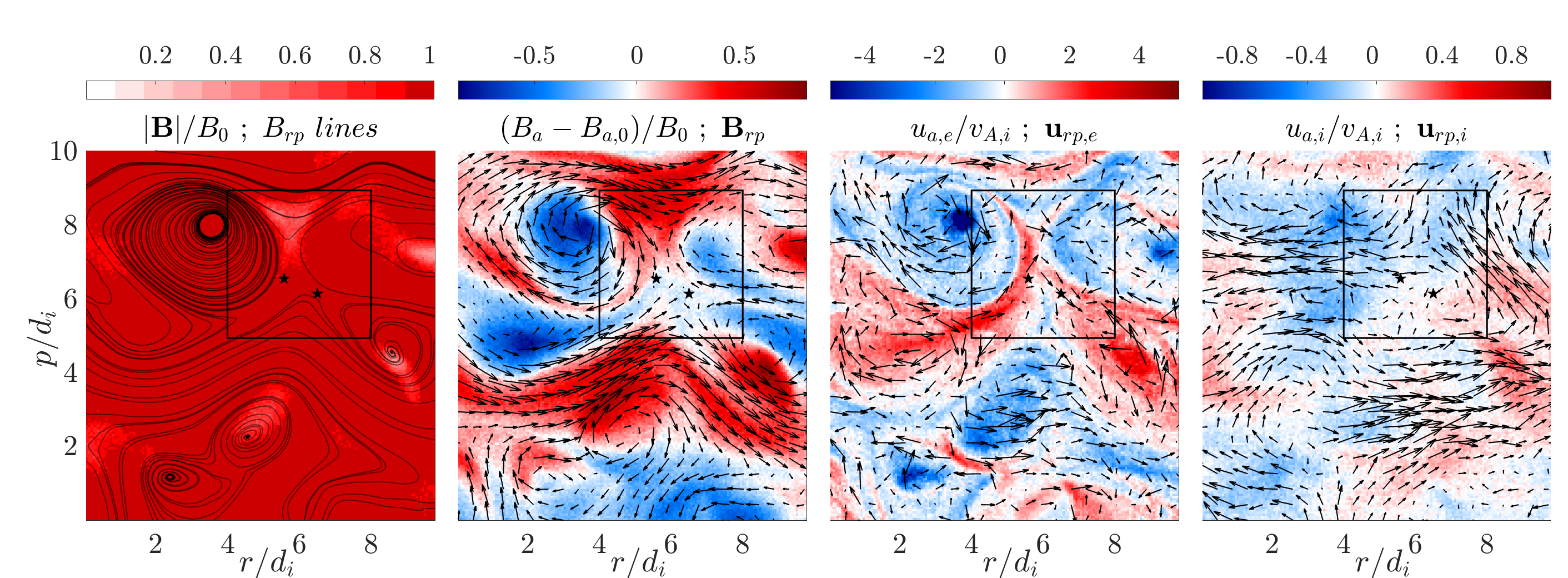}};
    \draw (-7.8, 3.0) node {a)};
    \draw (-3.6, 3.0) node {b)};
    \draw (0.7, 3.0) node {c)};
    \draw (4.9, 3.0) node {d)};
\end{tikzpicture}
\caption{2D cuts in the $rp$-plane at simulation time $t=120\omega_{pi}^{-1}$. a) Magnetic field magnitude $|\vec{B}|/B_{0}$. The black contours represent the in-plane magnetic-field lines and the black stars represent two x-points. b) Out of plane component magnetic field $(B_{a}-B_{a,0})/B_{0}$. The black arrows in this panel represent the in-plane magnetic vectors ($\vec{B}_{rp}$). c) Out of plane electron speed $u_{a,e}/v_{A,i}$. The black arrows in this panel represent the in-plane electron velocity vectors ($\vec{u}_{rp,e}/v_{A,i}$). d) out of plane ion speed $u_{a,i}/v_{A,i}$. The black arrows in this panel represent the in-plane ion velocity vectors ($\vec{u}_{rp,i}$). In all panels the black square outlines the diffusion region.}
\label{fig:Ba_uea_uia_2000}
\end{figure*}

{In the following analysis we use the RF ($r,p,a$) and select a cubic region of size 10 $d_{i}$. Although the event is three-dimensional and the magnetic field lines extend in three dimensions, we select a 2D cut of the cubic region in the $rp$-plane similar to the green square in panel c) of Figure \ref{fig:three_composition}. This 2D cut is representative of the reconnection event as we show in Section \ref{sec:Agirotropy}.} 
 
Panel a) in Figure \ref{fig:Ba_uea_uia_2000} shows the magnetic-field magnitude in the region of interest normalized to $B_{0}$. {The black contours represent the in-plane magnetic-field lines which we compute by creating an array of seed points placed on the vertices of a squared grid in the $rp$-plane. Then, we use the in-plane magnetic field vectors to create the streamlines. We propagate the numerical integration in both directions: forwards and backwards.} Panel b) in Figure \ref{fig:Ba_uea_uia_2000} {shows $(B_{a}-B_{a,0})/B_{0}$, where $B_{a}$ is the out-of-plane component of the magnetic field and $B_{a,0}=\vec{B}_{0}\cdot\uvec{a}$ is the projection of $B_{0}$ on the $a$-direction. We subtract $B_{a,0}$ to improve the visibility of the multipolar configuration of this component.} The black arrows in this panel represent the in-plane magnetic-field vectors $\vec{B}_{rp} = B_{r} \uvec{r} + B_{p} \uvec{p}$. In order for reconnection to occur, the in-plane components of the magnetic fields of reconnecting structures must have {different} directions. {The plotted in-plane magnetic-field lines suggest the presence of effective separatrices between regions of opposite $\vec{B}_{rp}$ within the black square.}

The in-plane magnetic-field lines in panel a) along with the direction of the in-plane magnetic-field vectors suggest the presence of two x-points which we mark with two black stars, one located at {$r=5.58d_{i}$} and $p= 6.6d_{i}$ and the other at {$r=6.5d_{i}$} and $p=6.2d_{i}$.
{We establish the position of the x-points by identifying the saddle points of the in-plane magnetic field.}
The magnetic configuration is complex, and the black square outlines the central region in which the reconnection occurs. {Within this region, the magnetic field is non-uniform. The sub-region where $|\vec{B}| \approx 0$ represents a null region.} From now on, we refer to the region enclosing the x-points as the diffusion region. Since transverse 2D cuts to 3D magnetic flux-ropes resemble the geometry of magnetic islands, we now refer to the magnetic-field lines which are quasi-circular in panel a) as magnetic islands. 

Panel c) shows the out of plane component of the electron velocity $u_{a,e}$ normalized to the ion Alfvén speed $v_{A,i}$. The red color indicates electrons moving out of the plane whereas the blue color indicates electrons moving into the plane. The black arrows of this panel represent the in-plane electron velocity vectors $\vec{u}_{rp,e} = u_{r,e} \uvec{r} + u_{p,e} \uvec{p}$. Within the region of interest, there are counter-streaming electrons following the separatrices. Likewise, we locate electrons streaming out of the plane through the diffusion region. Within the magnetic islands the electrons stream into the plane. In most of the magnetic islands, the electrons follow quasi-circular orbits due to their magnetization. However, in the magnetic island centered at $r=4.5d_{i}$ and $p=2.2d_{i}$, the electrons demagnetize and traverse into the magnetic island connecting with the stream of electrons at the edge of the magnetic island. 

Panel d) shows the out of plane component of the ion velocity $u_{a,i}$ normalized to $v_{A,i}$. The black arrows in this panel represent the in-plane ion velocity vectors $\vec{u}_{rp,i} = u_{r,i} \uvec{r} + u_{p,i} \uvec{p}$. Within the diffusion region the out of plane ion velocity is small which suggests that the ion motion is mostly constrained to the plane. The in-plane motion, however, is considerable and the ions move across the separatrices since they are demagnetized.

\subsection{Particle agyrotropy in the diffusion region}
\label{sec:Agirotropy}

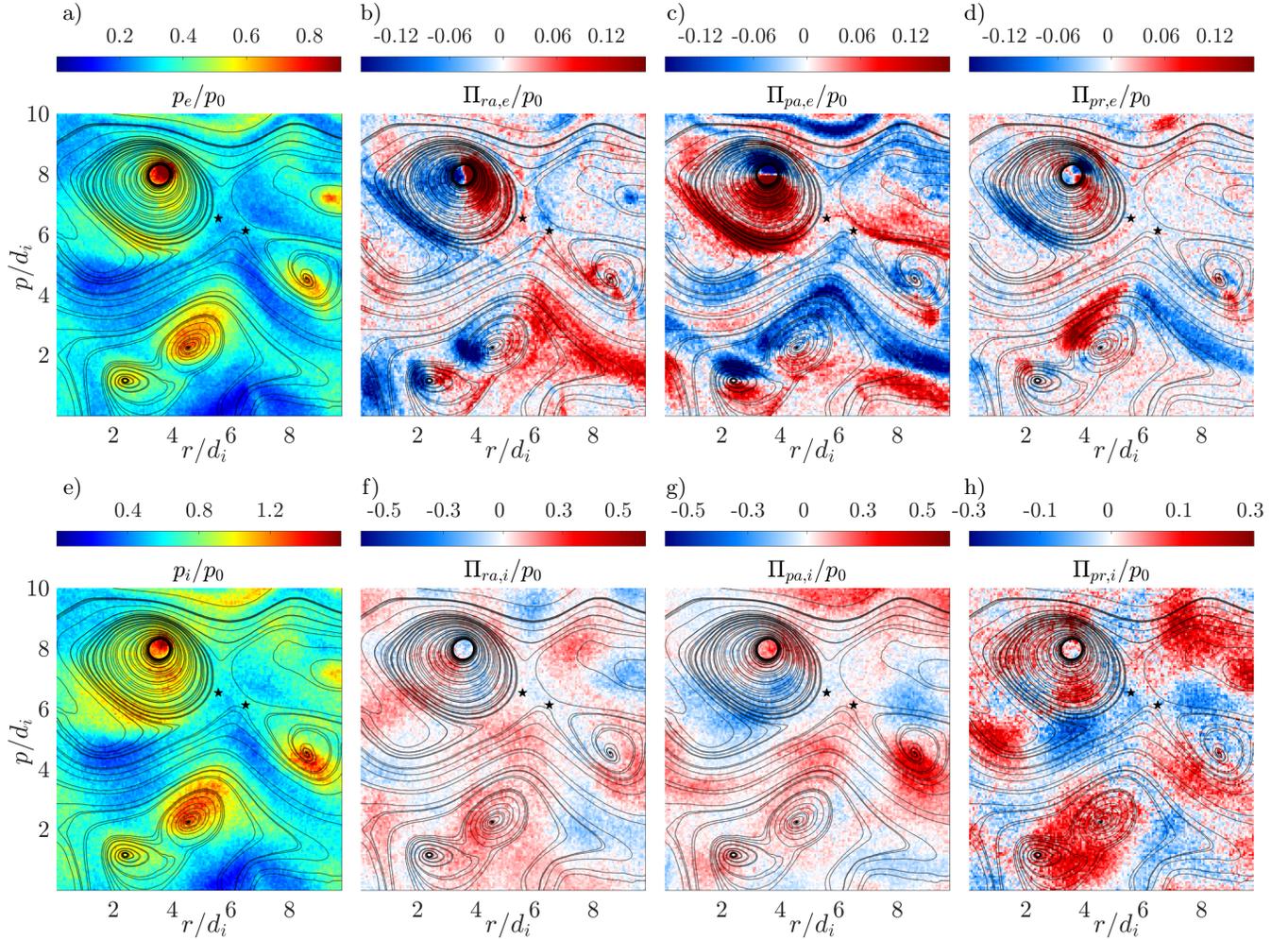
\begin{figure*}
\centering
\begin{tikzpicture}
    \draw (0, 0) node[inner sep=0] 
    {\includegraphics[width=1\linewidth]{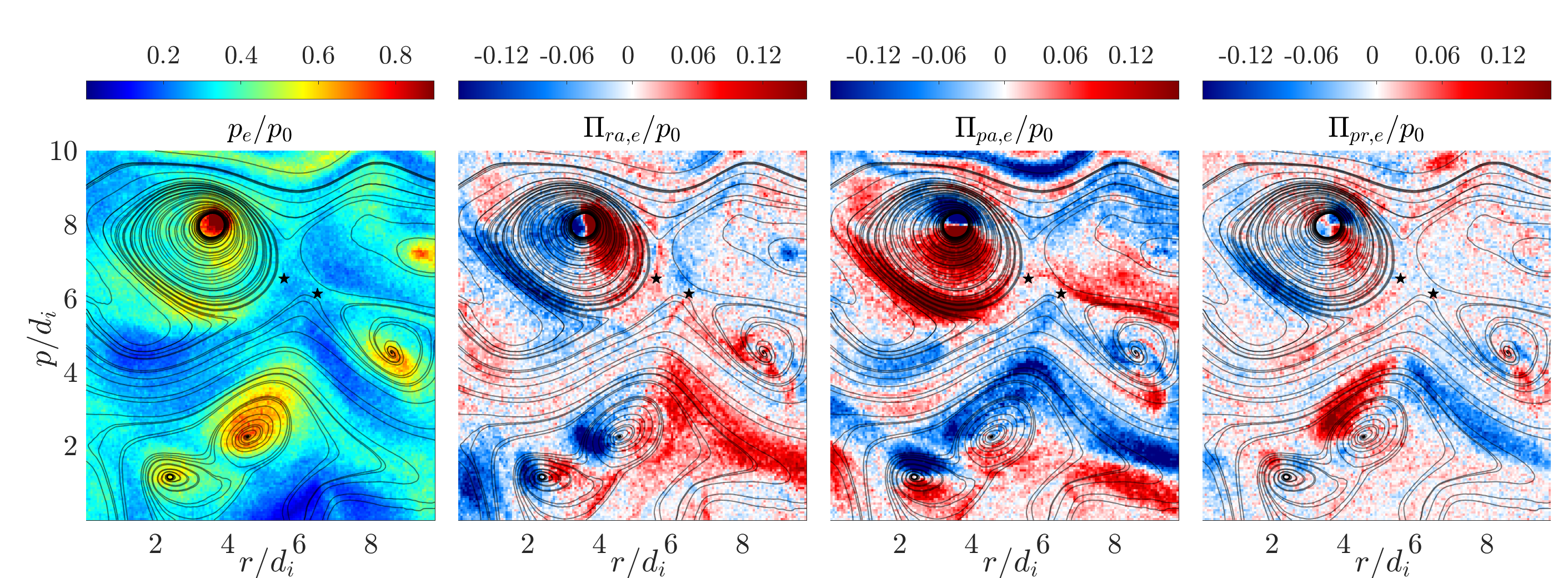}};
    \draw (-7.8, 3.0) node {a)};
    \draw (-3.6, 3.0) node {b)};
    \draw (0.7, 3.0) node {c)};
    \draw (4.9, 3.0) node {d)};
\end{tikzpicture}
\begin{tikzpicture}
    \draw (0, 0) node[inner sep=0] 
    {\includegraphics[width=1\linewidth]{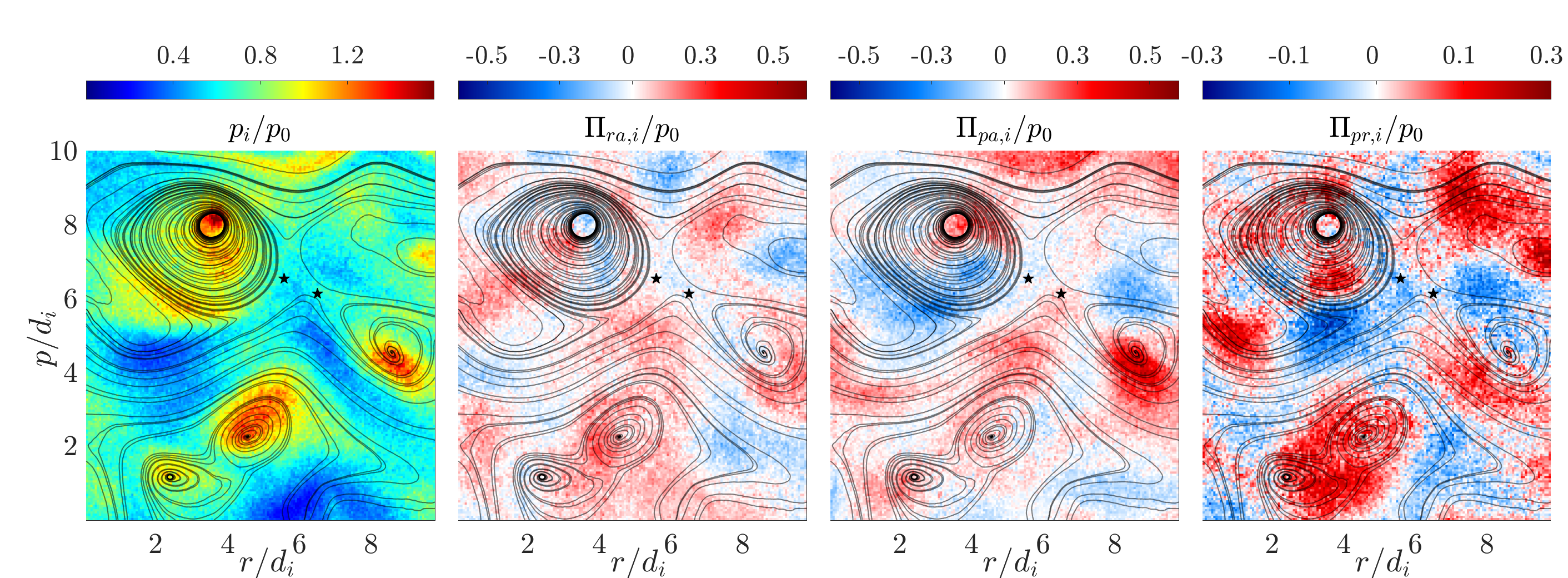}};
    \draw (-7.8, 3.0) node {e)};
    \draw (-3.6, 3.0) node {f)};
    \draw (0.7, 3.0) node {g)};
    \draw (4.9, 3.0) node {h)};
\end{tikzpicture}
\caption{2D cuts of the pressure tensor components in the $rp$-plane at the simulation time $t=120 \omega^{-1}_{pi}$. a) Electron scalar pressure $p_{e}/p_{0}$. Off-diagonal components of the electron pressure tensor: b) $\Pi_{ra,e}/p_{0}$, c) $\Pi_{pa,e}/p_{0}$  and d) $\Pi_{pr,e}/p_{0}$. e) Ion scalar pressure $p_{i}/p_{0}$. Off-diagonal components of the ion pressure tensor: f) $\Pi_{ra,i}/p_{0}$, g) $\Pi_{pa,i}/p_{0}$ and h) $\Pi_{ra,i}/p_{0}$.}
\label{fig:pepi_PI}
\end{figure*}

During the reconnection of magnetic flux-ropes, the plasma expansion/contraction is not isotropic. Therefore, at kinetic scales the plasma pressure of each species can develop anisotropy and agyrotropy. Figure \ref{fig:pepi_PI} shows our pressure terms according to Eqs. (\ref{eqn:pi_pressure_d}) and (\ref{eqn:PI_pressure_of}) for electrons and ions normalised to the initial ion pressure $p_{0}=n_{0}m_{i}v_{A,i}^{2}$. Panels a) and e) show the isotropic scalar pressure for electrons $p_{e}$ and ions $p_{i}$ respectively. For both species, the scalar pressure is greater inside the magnetic islands than outside due to the large density of particles (not shown here). Likewise, $p_{e}$ and $p_{i}$ display gradients along and across the separatrices. We find that $p_{e}$ is lower in the region between the magnetic islands as well as between the x-points compared to inside the magnetic islands. 

Panels b), c) and d) of Figure \ref{fig:pepi_PI} show the off-diagonal components of the electron pressure tensor according to Eq.~(\ref{eqn:PI_pressure_of}), $\Pi_{ra,e}$, $\Pi_{pa,e}$, and $\Pi_{pr,e}$, respectively. We here introduce our notation $\langle ... \rangle$ for the spatial average of a quantity over a given domain. The averages over the sub-domain of $|\Pi_{ra,e}|$,  
and $|\Pi_{pa,e}|$, $\langle |\Pi_{ra,e}| \rangle$,  
and $\langle |\Pi_{pa,e}| \rangle$, are about 10$\%$ of $\langle p_{e} \rangle$. $\Pi_{ra,e}$ and $\Pi_{pa,e}$ present a strong dipole-like configuration centered on the magnetic islands. There is a shallower, yet visible, gradient in $\Pi_{ra,e}$, $\Pi_{pa,e}$ and $\Pi_{pr,e}$ in the region between the islands as well as in the diffusion region. Conversely, $\Pi_{pr,e}$ exhibits a quadrupolar configuration within the magnetic islands. The non-zero values of $\Pi_{ra,e}$, $\Pi_{pa,e}$ and $\Pi_{pr,e}$ show that the plasma is agyrotropic, suggesting that small-scale kinetic processes occur. Similar patterns are reported along the separatrices of 2D collisionless reconnection \citep{yin2001hybrid,schmitz2006kinetic,wang2015comparison} and laminar 3D collisionless reconnection \citep{wang2018electron}. However, unlike previous studies, we observe the same patterns within the magnetic islands of turbulent 3D magnetic reconnection. This is a fundamental difference between the reconnection that occurs in turbulence and steady state reconnection that occurs in Harris current-sheet configurations. 

\begin{figure*}
\centering
\begin{tikzpicture}
    \draw (0, 0) node[inner sep=0] 
    {\includegraphics[width=1\linewidth]{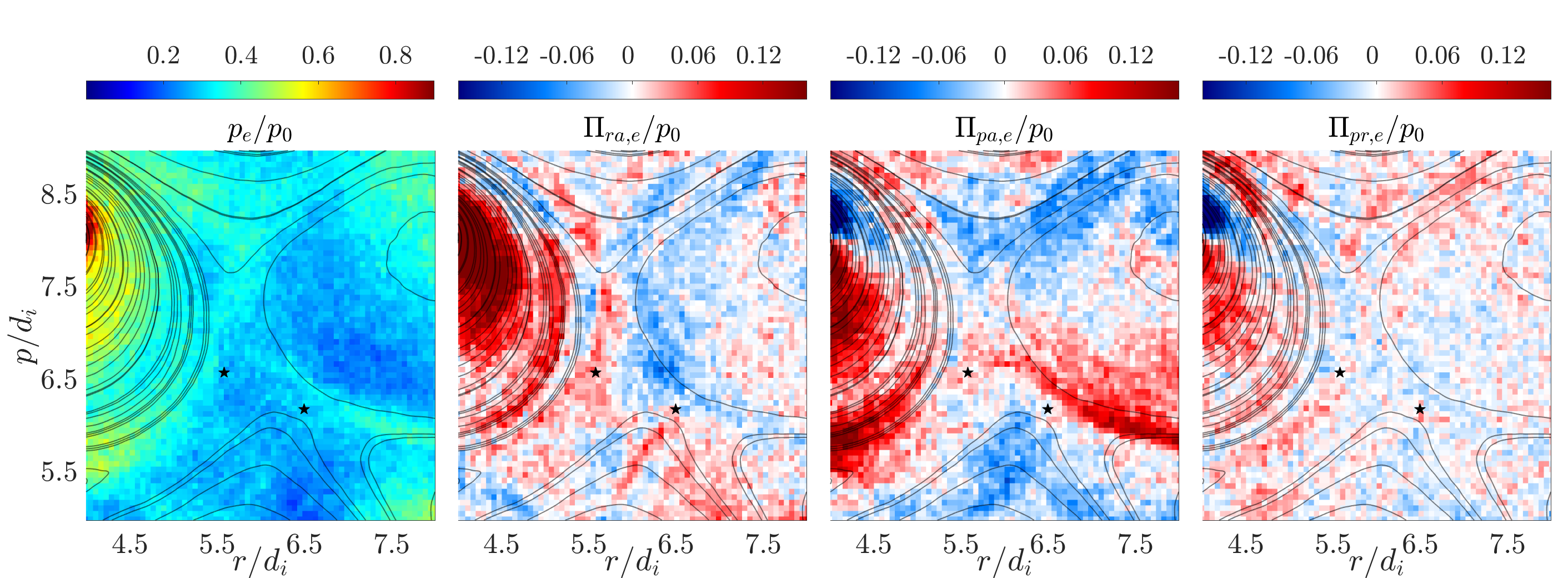}};
    \draw (-7.8, 3.0) node {a)};
    \draw (-3.6, 3.0) node {b)};
    \draw (0.7, 3.0) node {c)};
    \draw (4.9, 3.0) node {d)};
\end{tikzpicture}
\begin{tikzpicture}
    \draw (0, 0) node[inner sep=0] {\includegraphics[width=1\linewidth]{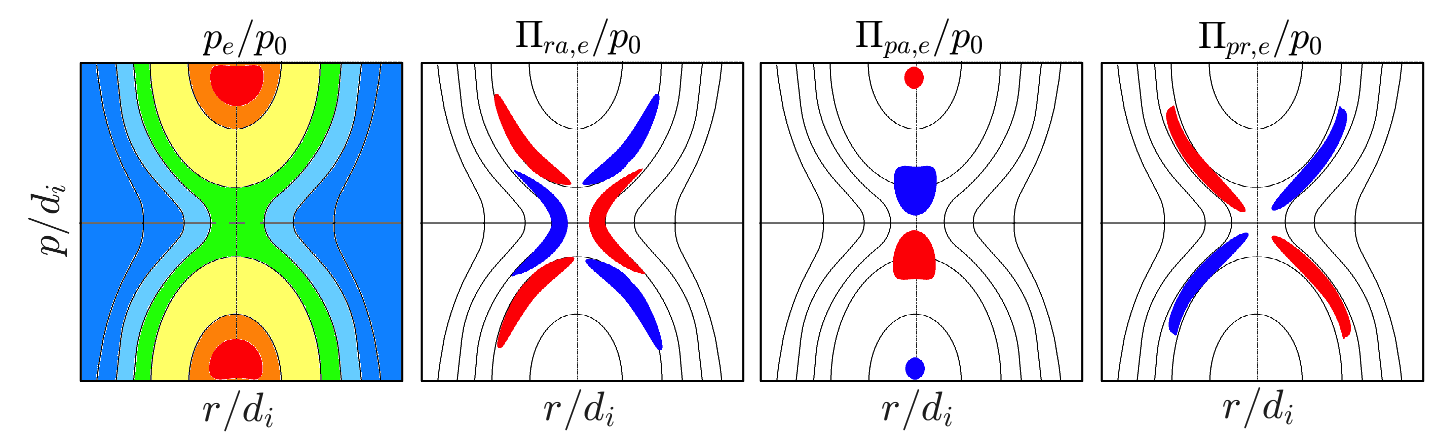}};
    \draw (-7.8, 2.5) node {e)};
    \draw (-3.6, 2.5) node {f)};
    \draw (0.7, 2.5) node {g)};
    \draw (4.9, 2.5) node {h)};
\end{tikzpicture}
\caption{Magnification of the region delimited by the black square in Figure \ref{fig:Ba_uea_uia_2000} in the $rp$-plane at the simulation time $t=120 \omega^{-1}_{pi}$. a) Electron scalar pressure $p_{e}$. Off-diagonal components of the electron pressure tensor: b) $\Pi_{ra,e}$, c) $\Pi_{pa,e}$  and d) $\Pi_{pr,e}$. Panel e) shows a sketch of the patterns of the scalar pressure emerging in 2D simulations of reconnection, and f) to h) show sketches of the off diagonal terms of the electron pressure based on 2D Harris current sheet reconnection without guide field \citep{yin2001hybrid}.}
\label{fig:pe_PI_12}
\end{figure*}

Panels f), g) and h) of Figure \ref{fig:pepi_PI} show the off-diagonal components of the ion pressure tensor according to Eq. (\ref{eqn:PI_pressure_of}), $\Pi_{ra,i}$, $\Pi_{pa,i}$ and $\Pi_{pr,i}$ respectively. The off-diagonal terms for ions, unlike electrons, have a less coherent pattern attached to the in-plane magnetic field topology. The reason for this detachment lies in the de-magnetization of the ions at these scales. Nevertheless, there is a gradient of these terms suggesting agyrotropy effects in the ion dynamics as well.     

Figure \ref{fig:pe_PI_12} shows a magnification of the region enclosed by the black square in Figure \ref{fig:Ba_uea_uia_2000}. Panels a) to d) show a magnification of the electron pressure terms shown in panels a) to d) of Figure \ref{fig:pepi_PI}. To make a direct comparison with previous 2D studies, panels e) to h) show sketches summarizing known patterns associated with the electron pressure components that emerge from 2D collisionless reconnection in the absence of a guide field \citep{yin2001hybrid,schmitz2006kinetic,wang2015comparison}. In this region, unlike within the magnetic islands of Figure \ref{fig:pepi_PI}, our simulation results of the electron pressure patterns match those patterns shown in the sketches in panels e) to h) in the region where the magnetic field has a local minimum according to panel a) in Figure \ref{fig:three_composition}. However, below the x-point located at {$r=5.58 d_{i}$} and $p=6.6d_{i}$, the pattern no longer corresponds to the sketched expectations. Moreover, $\Pi_{pr,e}$ is less coherent, and we do not recognize a clear quadrupolar configuration as in the sketches for the 2D case.   

\begin{figure*}
\centering
\begin{tikzpicture}
    \draw (0, 0) node[inner sep=0] 
    {\includegraphics[width=0.87\linewidth]{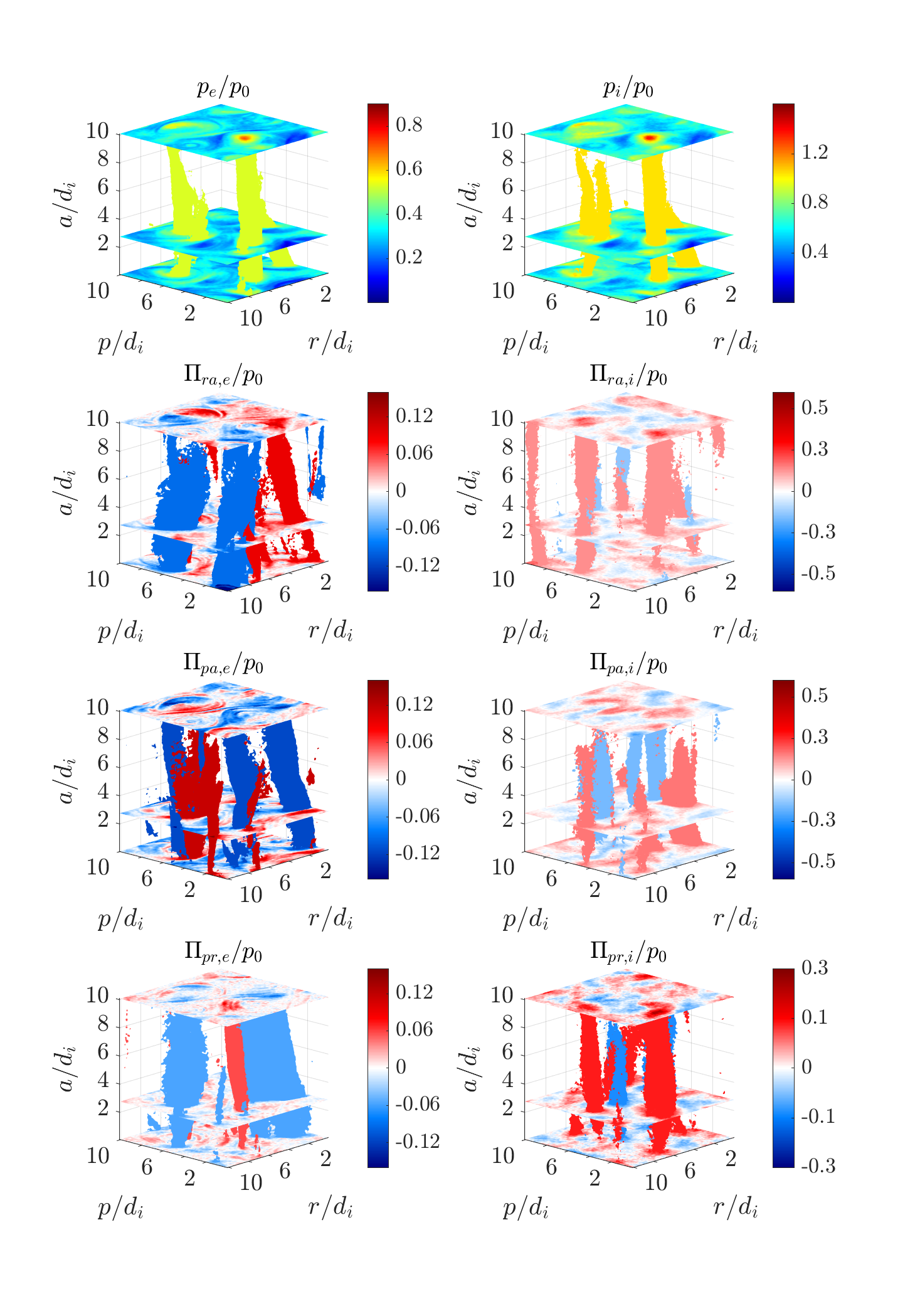}};
    \draw (-7.4, 10.1) node {a)};
    \draw (-7.4, 4.9) node {b)};
    \draw (-7.4, -0.2) node {c)};
    \draw (-7.4, -5.7) node {d)};
    \draw (0.2, 10.1) node {e)};
    \draw (0.2, 4.9) node {f)};
    \draw (0.2, -0.2) node {g)};    
    \draw (0.2, -5.7) node {h)};
\end{tikzpicture}
\caption{2D cuts in the $rp$-plane and isosurfaces of the pressure tensor components at the simulation time $t=120 \omega^{-1}_{pi}$. a) Electron scalar pressure $p_{e}/p_{0}$. Off-diagonal components of the electron pressure tensor: b) $\Pi_{ra,e}/p_{0}$, c) $\Pi_{pa,e}/p_{0}$,  and d) $\Pi_{pr,e}/p_{0}$. e) Ion scalar pressure $p_{i}/p_{0}$. Off-diagonal components of the ion pressure tensor: f) $\Pi_{ra,i}/p_{0}$, g) $\Pi_{pa,i}/p_{0}$, and h) $\Pi_{ra,i}/p_{0}$.}
\label{fig:pei_PI}
\end{figure*}
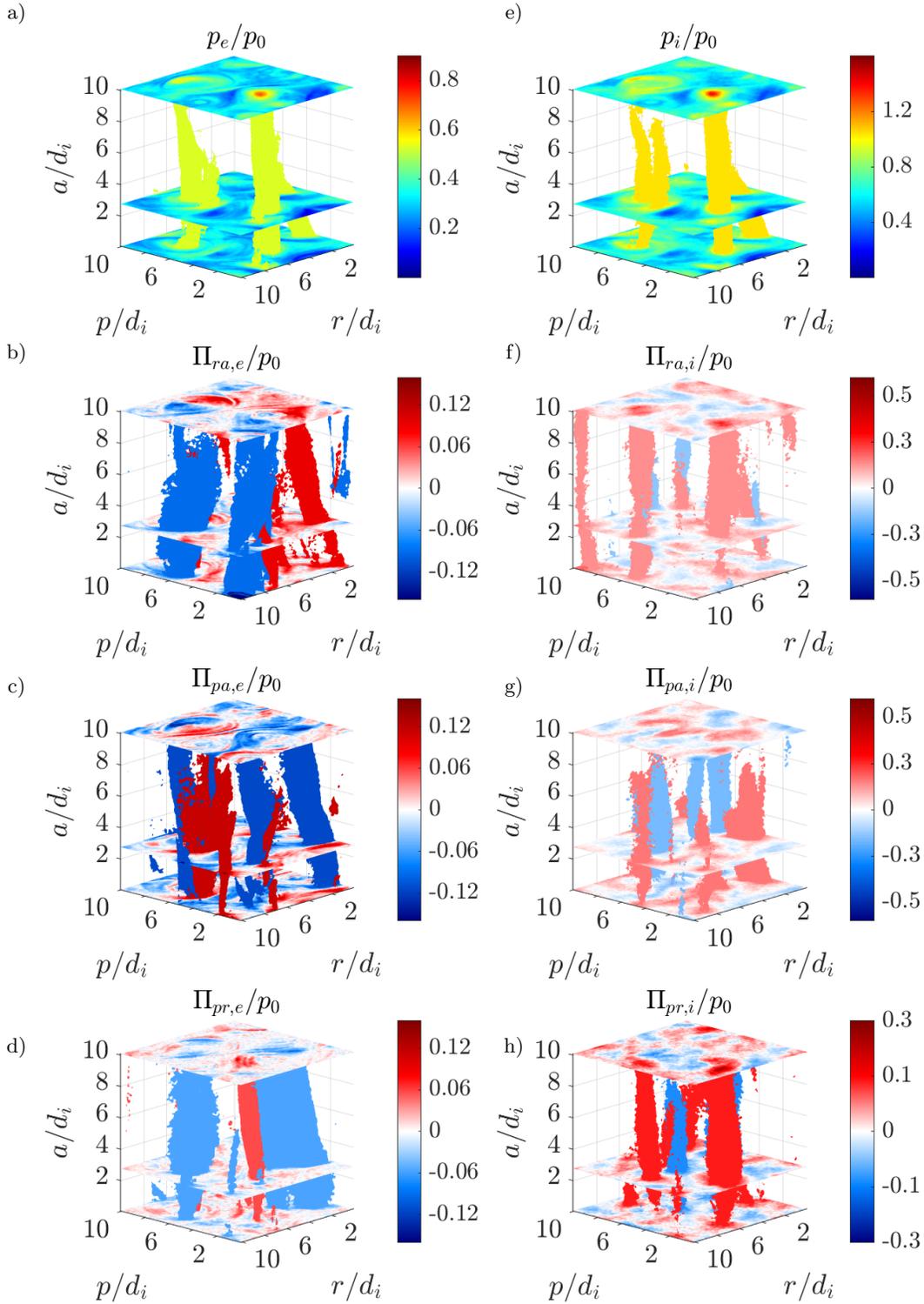

{Figure~\ref{fig:pei_PI} shows a 3D representation of the pressure components for electrons in  panels a) through d), and for ions in panels e) through h). The 2D cut at $a=2.76d_{i}$ corresponds to the 2D cut in Figure~\ref{fig:pepi_PI}. The plotted 3D structures are isosurfaces of the pressure component depicted on the 2D planes. For a given quantity $\psi$, the value of the isosurfaces corresponds to $S_{\psi} = \pm (\langle |\psi| \rangle + 2\sigma_{|\psi|})$, where $\sigma_{|\psi|}$ is the standard deviation of $|\psi|$. The isosurfaces in panels a) through h) have the shape of elongated and thin surfaces with local curvatures along the $a$-axis. The agyrotropic patterns in Figure~\ref{fig:pepi_PI} extend for $\sim 5d_{i}$ along the ${a}$-axis.}

\subsection{Energy transfer and transport}
\label{sec:energy_distri}

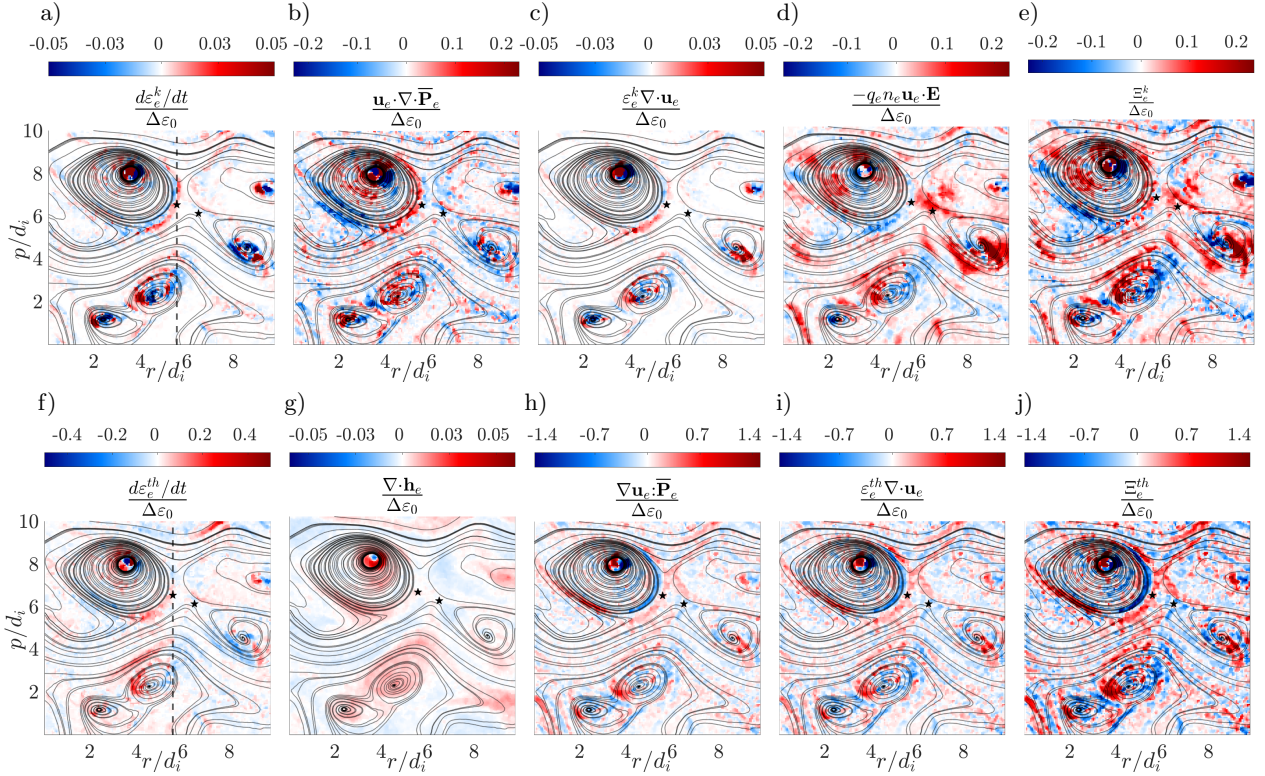
\begin{figure*}
\centering
\begin{tikzpicture}
    \draw (0, 0) node[inner sep=0] 
    {\includegraphics[width=0.95\linewidth]{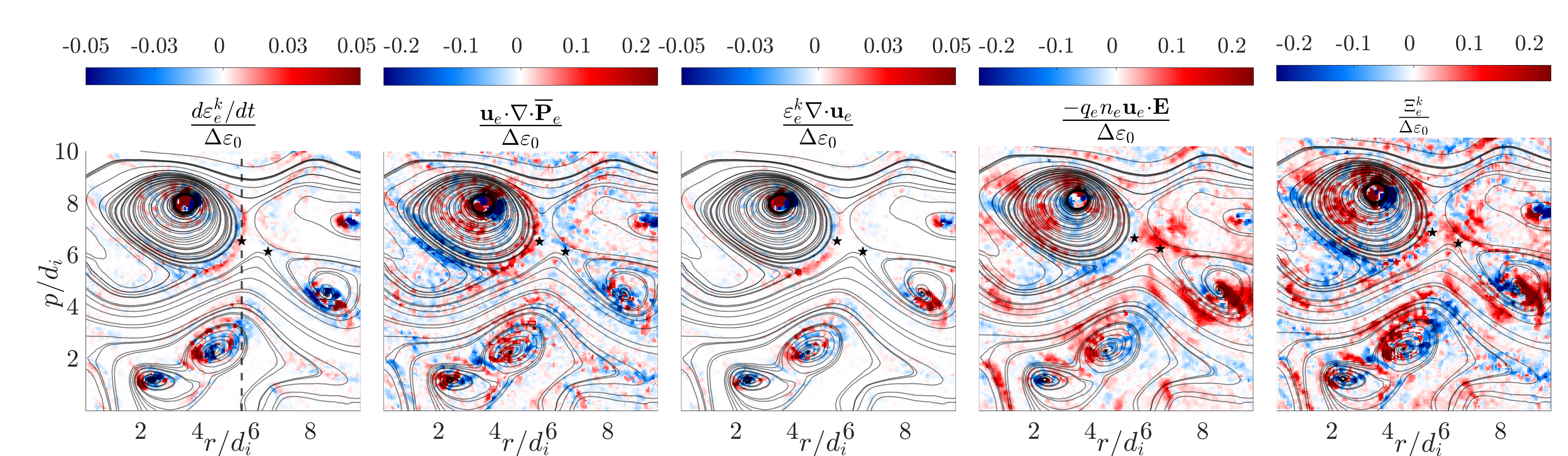}};
    \draw (-7.6, 2.4) node {a)};
    \draw (-4.3, 2.4) node {b)};
    \draw (-1.1, 2.4) node {c)};
    \draw (2.2, 2.4) node {d)};
    \draw (5.4, 2.4) node {e)};
\end{tikzpicture}
\begin{tikzpicture}
    \draw (0, 0) node[inner sep=0]     
    {\includegraphics[width=0.95\linewidth]{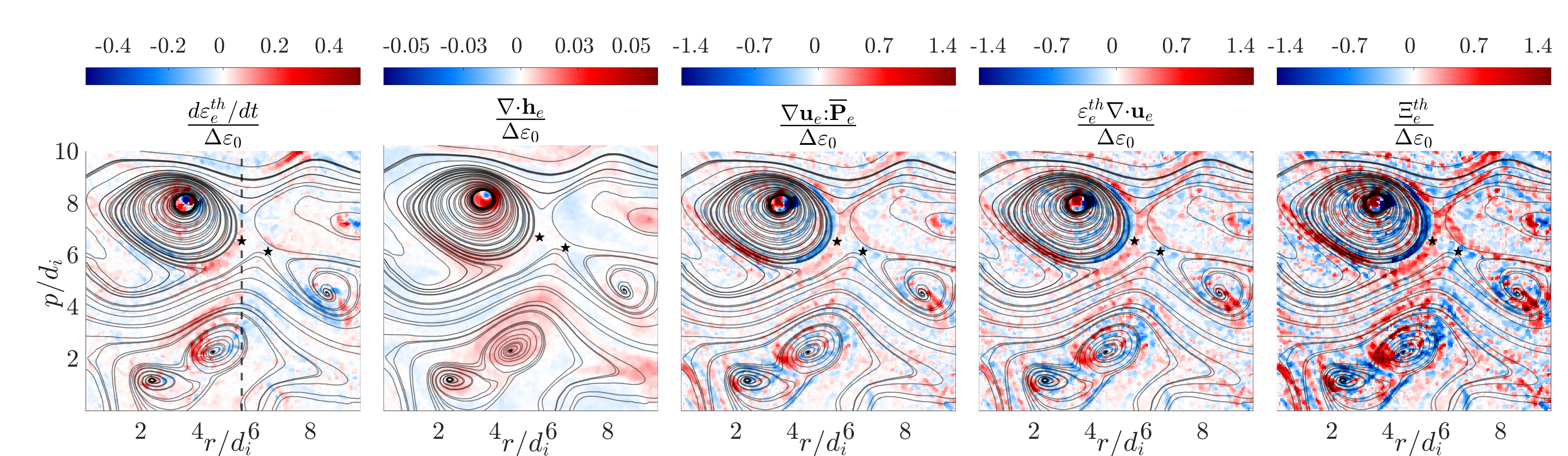}};
    \draw (-7.6, 2.4) node {f)};
    \draw (-4.3, 2.4) node {g)};
    \draw (-1.1, 2.4) node {h)};
    \draw (2.2, 2.4) node {i)};
    \draw (5.4, 2.4) node {j)};
\end{tikzpicture}
\caption{2D cuts in the $rp$-plane at the simulation time $t=120 \omega^{-1}_{pi}$. Panels a) to e): kinetic power density terms for electrons. Panels f) to j): thermal power density terms for electrons. All quantities are normalised to $\Delta \varepsilon_{0}=\omega_{pi}m_{i}v_{A,i}^{2}$. {The vertical dashed lines in panels a) and f) show the 1D trajectory for our 1D analysis.}} 
\label{fig:power_e}
\end{figure*}

We use the power density expressions for the kinetic energy in Eq. (\ref{eqn:firstmomener_text}) and thermal energy in Eq. (\ref{eqn:secondmomener_text}) to describe the energy transfer and transport associated with our reconnection event. To compute the partial time derivatives of a quantity, we use a central difference approach. Since the Alfvén transient time is $\sim 100$ $\omega^{-1}_{pi}$, a time resolution of $6$ $\omega_{pi}^{-1}$ is sufficient to capture the relevant dynamics of interest. To estimate the spatial derivatives, we use a standard cell-centered first neighbors approach. We calculate all scalar products cell-wise in the simulation domain. Panels a) to e) of Figure \ref{fig:power_e} show 2D cuts of each term in Eq. (\ref{eqn:firstmomener_text}) for electrons normalized to $\Delta \varepsilon_{0}=\omega_{pi}m_{i}v_{A,i}^{2}$. 

Panel a) shows, at the simulation time $t=120 \omega^{-1}_{pi}$, the total time derivative of the kinetic energy density ${d\varepsilon^{k}_{e}}/{dt}$. The domain exhibits considerable temporal changes of the kinetic energy density at the centers of the magnetic islands. We also detect negative ${d\varepsilon^{k}_{e}}/{dt}$ at the edge of the top left magnetic island and positive ${d\varepsilon^{k}_{e}}/{dt}$ in the diffusion region. Conversely, there is almost no change in $\varepsilon^{th}_{e}$ in the region between the x-points. 

Panel b) shows the scalar product $\vec{u}_{e}\cdot \nabla \cdot \overline{\vec{P}}_{e}$ which quantifies the change of kinetic energy due to the advection of the pressure tensor. This energy change is transported by the electron flow. The quantity $\vec{u}_{e}\cdot \nabla \cdot \overline{\vec{P}}_{e}$ is also known as the pressure work \citep{fadanelli2021energy}. There is a strong conversion of energy associated with the pressure work at the center of the magnetic islands. However, the energy change associated with this term is around 10 times greater than the local $d\varepsilon^{k}_{e}/dt$. Unlike $d\varepsilon^{k}_{e}/dt$, at the edge of the top left magnetic island, there is a strong gradient of $\vec{u}_{e}\cdot \nabla \cdot \overline{\vec{P}}_{e}$ from the left-hand side of the magnetic island to the right-hand side. In addition, $\vec{u}_{e}\cdot \nabla \cdot \overline{\vec{P}}_{e}$ has a local minimum in the region between the x-points. 

Panel c) shows $\varepsilon^{k}_{e}\nabla\cdot\vec{u}_{e}$ which represents the kinetic energy change due to divergent or convergent flow patterns in the electron bulk velocity. Like for the previous terms, $\varepsilon^{k}_{e}\nabla\cdot\vec{u}_{e}$ is greater at the center of the magnetic islands than in the region between them. There is no noticeable gradient of this terms between the x-points. Although panels a), b) and c) show similar patterns in their signs, there are local differences, specially in the diffusion region. 

Panel d) shows $-q_{e}n_{e}(\vec{u}_{e}\cdot\vec{E})$ which represents the energy exchange between the electrons and the electric field. We find a considerable energy conversion not only within the magnetic islands but also in the region between the islands as well as in the region between the x-points. In the region between the x-points, the electrons gain kinetic energy from the electric field. Along the separatrix next to the top left island, the electron bulk motion is decelerated by the electric field. Comparing panels b) and d) $\vec{u}_{e}\cdot \nabla \cdot \overline{\vec{P}}_{e}$ and $-q_{e}n_{e}(\vec{u}_{e}\cdot\vec{E})$ balance with each other in the diffusion region.

Panel e), shows $\Xi^{k}_{s}$, which we compute as the sum of all terms at the left hand side of Eq. (\ref{eqn:firstmomener_text}). There are regions with positive and negative $\Xi^{k}_{e}$ within the magnetic islands. On the contrary, $\Xi^{k}_{e}$ is predominately positive within the diffusion region and along the separatrices. 

Although we do not include binary collisions in our code explicitly, we acknowledge that the finite number of macro particles affects the system in a way similar to collisions, and leads to an undersampling of non-thermal fine structure in the velocity distribution function, which generates a loss of information and thus increase in entropy. 

In a real plasma, binary collisions between particles drive the system to a thermal equilibrium, thus smoothing out the distribution function. Similarly, a finite number of particles represents a low number of counts to compute the statistical measures. Therefore, when computing macroscopic quantities, the contribution from non-thermal particles is overshadowed by the core distribution. It is effectively a coarse-grained effect, similar to the actual effect of collisions, albeit on a different time scale. Although this effect occurs earlier in PIC simulations with a finite number of particles than in the real solar wind. We conjecture that the impact is ultimately comparable.

Panels f) to j) of Figure \ref{fig:power_e} show 2D cuts of each term in Eq. (\ref{eqn:secondmomener_text}) normalized to $\Delta \varepsilon_{0}$. Panel f) depicts $d\varepsilon^{th}_{e}/dt$. As in the kinetic-energy case, $d\varepsilon^{th}_{e}/dt$ has local extrema associated with the magnetic islands. The main change in $d\varepsilon^{th}_{e}/dt$ is due to the advective term $(\vec{u}_{e}\cdot \nabla) \varepsilon^{th}_{e}$. By direct comparison with panel b), we note similar power density patterns between $d\varepsilon^{th}_{e}/dt$ and $\vec{u}_{e}\cdot \nabla \cdot \overline{\vec{P}}_{e}$. 

For the heat-flux term $\nabla \cdot \vec{h}_{e}$, we do not directly compute $\nabla \cdot \vec{h}_{e}$ as a particle moment, but use a Hammett-Perkins approach \citep{hammett1990fluid} to estimate its contribution. This approach has been successfully applied in previous collisionless reconnection studies \citep{wang2015comparison,ng2015island,ng2017simulations}. In this framework, we estimate 
\begin{equation}
\nabla \cdot \vec{h}_{e} \approx
v^{th}_{e} \frac{1}{2} |k_{0}|Tr\left[P_{ij,e} - \langle P_{ij,e}\rangle - \delta_{ij} (n_{e} - \langle n_{e} \rangle)\langle {T}_{e}\rangle\right],
\label{eqn:divh_HP}
\end{equation}
where $v^{th}_{e} = \sqrt{2 k_{B}T_{e}/m_{e}}$ is the thermal speed of the electrons. The wave number $k_{0} = \sqrt{3}/|L_{s}|$ is a representative wave number associated with a sub-domain of volume $V_{s}=L_{s}^{3}$ where $L_{s}=10.08$ $d_{i}$ which we select as the region to study the energy conversion during the reconnection event. Panel g) shows our estimation of $\nabla \cdot \vec{h}_{e}$. There is a positive power density contribution from particle heatflux inside the magnetic islands. Conversely, there is a negative contribution in the regions between the magnetic islands. 

Panel h) depicts the energy transfer $\nabla \vec{u}_{e}:\tensndd{P}_{e}$ between kinetic and thermal energies. This term has contributions from the diagonal elements of the tensors which are associated with the isotropic energy transport and from the off-diagonal elements that quantify the agyrotropy in the plasma. There is positive $\nabla \vec{u}_{e}:\nabla \tensndd{P}_{e}$ in the region between the magnetic islands which is associated with counter streaming electrons. We locate an x-like structure centered in the region where the magnetic field strength exhibits a local minimum. In the region between the x-points as well as to the left of the diffusion region, $\nabla \vec{u}_{e}:\overline{\vec{P}}_{e}<0$.  

\makeatletter\onecolumngrid@push\makeatother
\clearpage
\makeatletter\onecolumngrid@pop\makeatother

Panel i) shows the thermal energy transport $\varepsilon^{th}_{e} \nabla \cdot \vec{u}_{e}$ associated with the compression/expansion of the electron flow. At first glance the positive/negative patterns in $\varepsilon^{th}_{e} \nabla \cdot \vec{u}_{e}$ seem very similar to the patterns in $\nabla \vec{u}_{e}:\overline{\vec{P}}_{e}$. The reason for this similarity is that the main energy transport in $\nabla \vec{u}_{e}:\overline{\vec{P}}_{e}$ is associated with the contribution of diagonal elements as we show in section \ref{sec:about_dissi}. However, we find local differences due to the agyrotropic contributions. From all terms on the left-hand sides of Eqs. (\ref{eqn:firstmomener_text}) and (\ref{eqn:secondmomener_text}), only the terms associated with the strain tensor present an extended asymmetric x-point-like structure in the diffusion region. Comparing panels c) and i), $\varepsilon^{th}_{e}\nabla \cdot \vec{u}_{e}$ is on average larger and forms broader structures than $\varepsilon^{k}_{e}\nabla \cdot \vec{u}_{e}$.

Panel j), shows $\Xi^{th}_{s}$, which we compute as the sum of all terms on the left hand side of Eq. (\ref{eqn:secondmomener_text}).This energy transfer is significant as the different terms on the left-hand side of Eq. (\ref{eqn:secondmomener_text}) do not sum to zero. 

In Figure \ref{fig:power_ei_1D}, we show vertical 1D cuts of the power density terms along the $p$-direction at $r = 5.58$ $d_{i}$ to visualize the relation between the different terms for plasma electrons. We further show a magnification of the region delimited by the black square from Figure \ref{fig:Ba_uea_uia_2000}. Panel a) shows the kinetic power density terms in Eq. (\ref{eqn:firstmomener_text}). We observe that the fluctuations in $d\varepsilon^{k}_{e}/dt$ (green line) and $\varepsilon^{k}_{e}\nabla \cdot \vec{u}_{e}$ (black line) are negligible compared with $\vec{u}_{e}\cdot\nabla\cdot\overline{\vec{P}}_{e}$ (red line) and $-q_{e}n_{e}\vec{E}\cdot\vec{u}_{e}$ (yellow line). However, there is a noticeable disturbance in all quantities in the range $p = 6.96$ $d_{i}$ to $p = 7.84$ $d_{i}$ which is located in the region of the diffusion region at which the magnetic field is nearly zero. Along the 1D cut, $\vec{u}_{e}\cdot\nabla\cdot\overline{\vec{P}}_{e}$ and $-q_{e}n_{e}\vec{E}\cdot\vec{u}_{e}$ are anti-correlated. This anti-correlation breaks when the disturbances in $d\varepsilon^{k}_{e}/dt$ and $\varepsilon^{k}_{e}\nabla \cdot \vec{u}_{e}$ occur. For this panel, the curve of $\Xi^{k}_{s}$ (blue line) changes sign when crossing the x-point.

Panel b) shows the thermal power density terms in Eq. (\ref{eqn:secondmomener_text}). Comparing panels a) and b), we observe that the fluctuations in the thermal power density terms are more pronounced than those in the kinetic power density. In panel b), the fluctuations in $d\varepsilon^{th}_{e}/dt$ (green line) and $\nabla \cdot \vec{h}_{e}$ (red line) are negligible compared with $\nabla \vec{u}_{s}:\overline{\vec{P}}_{e}$ (black line) and $\varepsilon^{th}_{e}\nabla \cdot \vec{u}_{e}$ (yellow line). Unlike in the kinetic power density case, the contributions from all terms in Eq. (\ref{eqn:secondmomener_text}) are either positive or negative at the same location showing no anti-correlation between the dominant terms. We note that $\Xi^{th}_{e}$ (blue line), unlike $\Xi^{k}_{e}$, is positive on both sides of the x-point. 

{Figure \ref{fig:power3d} shows a 3D representation of the kinetic power density terms in panels a) through d), and of thermal power density terms in panels e) through h). The 2D cut at $a=2.76d_{i}$ corresponds to the 2D cuts in Figure~\ref{fig:power_e}. The plotted 3D structures are isosurfaces of the power density terms depicted on the 2D planes. Panels a) through d) show that the isosurfaces of ${d\varepsilon^{k}_{e}}/{dt}, \vec{u}_{e}\cdot\nabla\cdot \overline{\vec{P}}_{e}$, and $\varepsilon^{k}_{e}$ are mostly thin filaments, whereas the isosurfaces of $-q_{e}n_{e}\vec{u}_{e}\cdot\vec{E}$ consist of broad patches. Moreover, there are more regions with $-q_{e}n_{e}\vec{u}_{e}\cdot\vec{E}>0$ than with $-q_{e}n_{e}\vec{u}_{e}\cdot\vec{E}<0$. Panels e) and f) show that the isosurfaces of ${d\varepsilon^{th}_{e}}/{dt}$ and $\nabla \cdot \vec{h}_{e}$ are also filamentary. Conversely, panels g) and h) show that the isosurfaces of $\vec{u}_{s}:\overline{\vec{P}}_{e}$ and $\varepsilon^{th}_{e}\nabla \cdot \vec{u}_{e}$ are mainly thin sheets.}

{Figure \ref{fig:xikinther_e} depicts isosurfaces of $\Xi^{k}_{e}$ in panel a) and $\Xi^{th}_{e}$ in panel b). The most evident isosurface of $\Xi^{k}_{e}$ is a filament located within the reconnecting flux-rope. Conversely, the isosurfaces of $\Xi^{th}_{e}$ are mostly thin sheets connected to the flux ropes.}




\subsection{Time evolution}

PIC simulations are affected by finite particle size, finite number of particles, and numerical integration errors that are effectively ``collisional'' contributions since they generate phase-space particle diffusion  \citep{hockney1971measurements,dawson1983particle,klimontovich2013statistical,birdsall2018plasma,grovselj2019fully}. Although the right-hand sides of the power-density relations in Eqs.~(\ref{eqn:firstmomener_text}) and (\ref{eqn:secondmomener_text}) include the contribution from numerical sources such as round-off errors and numerical heating, they also include contributions from the averaged, secular (including quasi-linear) correlations between fields and the particle distribution functions \citep{klein2016measuring,howes2017diagnosing,klein2017diagnosing}. As shown by the field--particle correlation method, meaningful averages of the non-linear correlations between the fluctuating electric field and the fluctuating perturbation of the distribution function define the secular transfer of energy from the fields to the particles. Thus, even in a purely collisionless plasma, the right hand sides of Eqs.~(\ref{eqn:firstmomener_text}) and (\ref{eqn:secondmomener_text}), after suitable averaging, is not exactly zero. In this interpretation, the averaging over higher-order field-particle correlations introduces the irreversibility and thus the dissipation into the kinetic description. All PIC systems share this behaviour with statistical particle systems in reality.

In this section, we present a time-evolution analysis of the energy-density terms in order to estimate the nature of $\Xi^{k}_{s}$ and $\Xi^{th}_{s}$. Figure~\ref{fig:time_evolution}a) shows the time evolution of the energy densities averaged over the full simulation domain (solid curves, subscript $full$) and averaged over the sub-domain (dashed curves, subscript $sub$). The curves are normalized to $\varepsilon_{0}=m_{i}v_{A,i}^{2}$. The total energy density is $\varepsilon^{T}=\varepsilon^{k}_{e}+\varepsilon^{k}_{i}+\varepsilon^{th}_{e}+\varepsilon^{th}_{i}+\varepsilon^{em}$. The averaged total energy densities $\langle \varepsilon^{T}_{full}\rangle$ and $\langle \varepsilon^{T}_{sub}\rangle$ (green curves) remain approximately constant. This suggests that  numerical heating is negligible in our energy balance.

\begin{figure*}
\centering
\begin{tikzpicture}
\draw (0, 0) node[inner sep=0] {\includegraphics[width=1\linewidth]{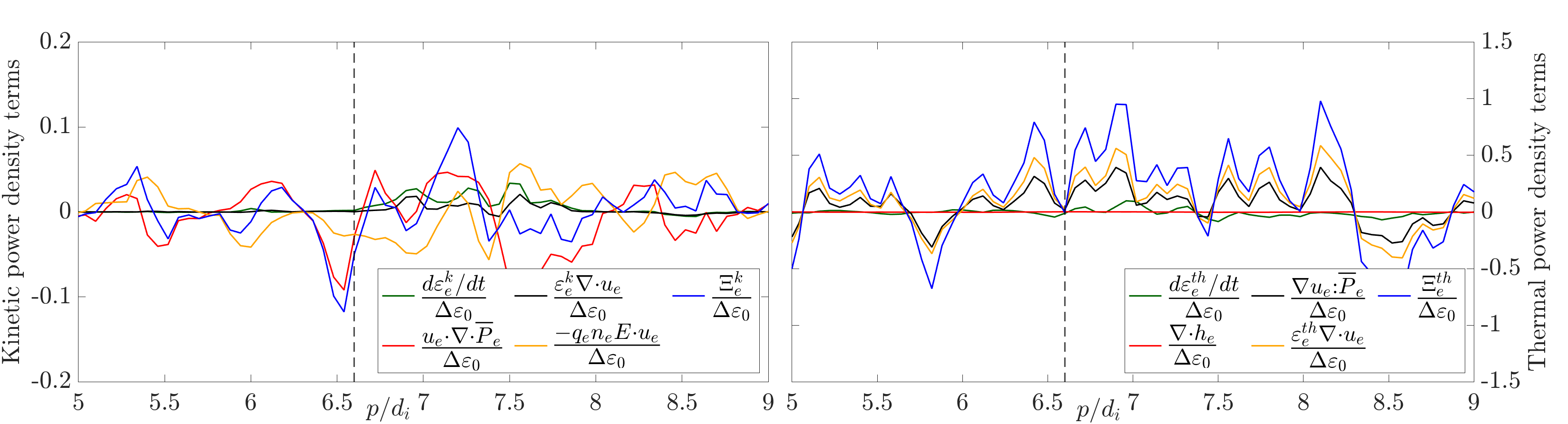}};
\draw (-7.8, 2.4) node {a)};
\draw (0.3, 2.4) node {b)};
\end{tikzpicture}
\caption{1D cuts of the power density terms along the $\hat{p}$-direction at $r = 5.58$ $d_{i}$ and at the simulation time $t=120 \omega^{-1}_{pi}$. a) Kinetic power density terms in Eq. (\ref{eqn:firstmomener_text}). b) Thermal power density terms in Eq. (\ref{eqn:secondmomener_text}). The vertical dashed line represents the crossing of the x-point {$r=5.58 d_{i}$} and $p=6.6 d_{i}$.}
\label{fig:power_ei_1D}
\end{figure*}

The thermal energy densities of both the ions (black curves) and  the electrons (magenta curves) are greater than the kinetic energy densities of both the ions (yellow curves) and electrons (red curves). When averaged over the sub-domain, the energy densities present more variability due to the in-flowing and out-flowing of energy density through the boundaries of the sub-domain. Nevertheless, the time evolution of the quantities $\varepsilon_{full}$ and $\varepsilon_{sub}$ is approximately comparable.     

Figure \ref{fig:time_evolution}b) depicts the time evolution of the absolute values of the energy-density rates $\Delta \langle \varepsilon \rangle / \Delta t$ (dashed curves) and the dissipative power-density rates $\Xi$ (solid-dotted curves), now averaged over the full simulation domain and normalized to $\varepsilon^{T}_{full}$. The time difference $\Delta t = 6\omega^{-1}_{pi}$ is the difference between two consecutive output times of our simulation. 

As shown in panel b), in the case of the ions, the thermal energy-density rate (black-dashed curve) and the kinetic energy-density rate (yellow-dashed curve) are greater than the dissipative power-density terms $\Xi^{th}_{i}$ (black solid-dotted curve) and $\Xi^{k}_{i}$ (yellow solid-dotted curve). The same ordering applies to the electron case in which the thermal energy-density rate (magenta dashed curve) and kinetic energy-density rate (red dashed curve) are greater than $\Xi^{k}_{e}$ (red solid-dotted curve) and $\Xi^{k}_{e}$ (magenta solid-dotted curve). For both species, $\Delta \varepsilon^{th}$ increases faster than $\Delta \varepsilon^{k}$. 

During the initial phase of the simulation ($t\omega_{pi} \lesssim 100$), we find that $\langle \Xi^{k}_{e}\rangle > \langle \Xi^{th}_{e} \rangle$ when averaged over the full simulation domain. Afterwards, for $t\omega_{pi} \gtrsim 100$, we find that $\langle \Xi^{k}_{e}\rangle < \langle \Xi^{th}_{e} \rangle$ when averaged over the full simulation domain until the simulation ends. The time $t\omega_{pi} \approx 100$ corresponds to the moment at which the overall $J^{rms}$ reaches its global maximum in our simulation and significant magnetic reconnection sets in \citep{agudelo2021three}.

The total energy-density rate $\Delta \varepsilon^{T}_{full}$ (green dashed curve) is lower than $\Delta \varepsilon^{k}$ and $\varepsilon^{th}$ for both species.  Moreover, $\langle \Xi^{th}_{e} \rangle$ and $\langle \Xi^{k}_{e} \rangle$ are negligible compared with the kinetic and thermal energy-density rates. This suggests that any irreversible energy transfer and thus numerical heating are negligible for the energy balance in our simulation. However, $\Xi^{th}_{e}$ and $\Xi^{k}_{e}$ are locally important near the reconnection region, see Figure \ref{fig:power_ei_1D}.

\subsection{Comparison with damping and heating proxies}
\label{sec:about_dissi}

In recent studies \citep{yang2017energy,pezzi2019energy,matthaeus2020pathways,pezzi2021dissipation} the collisionless energy dissipation problem is tackled by studying quantities such as the Zenitani parameter defined in Eq. (\ref{eqn:zenitani}) and the strain-pressure interaction defined in Eq. (\ref{eqn:double_duP}). We also explore these damping and heating proxies for comparison with our methods. Figure \ref{fig:D_pepi_PI_22} depicts 2D cuts in the $rp$-plane and 1D cuts of these damping and heating proxies. Panel a) shows $D_{ze}$. Similar to our kinetic and thermal power density terms, the magnetic islands present strong variations of $D_{ze}$. On the contrary, in the diffusion region, we see a coherent positive $D_{ze}$ signature. 

Panel b) shows $p\theta_{e}$. The positive/negative patterns of this quantity are almost identical to our patterns of $\nabla \vec{u}_{e}:\overline{\vec{P}}_{s}$ (panel h) in Figure \ref{fig:power_e}. This similarity illustrates that the main contribution to the strain-tensor interaction comes from the diagonal elements of the strain tensor. 

Panel c) shows $Pi \mh D_{e}$. Although the positive/negative patterns in $Pi \mh D_{e}$ are similar to those in $p\theta_{e}$, $Pi \mh D_{e}$ presents clear differences, especially near the null region where $Pi \mh D_{e}$ has the opposite sign of $p\theta_{e}$ along the separatrices. Moreover, along the separatrices, $|Pi$-$D_{e}| > D_{ze}$ and they share the same sign whereas in the region between the x-points $Pi \mh D_{e} < 0$ and $D_{ze}>0$.

Panel d) shows 1D cuts of $D_{ze}$ (blue line), $p\theta_{e}$ (red line) and $Pi \mh D_{e}$ (black line). We find that $p\theta_{e}$ is highly variable and, on average, greater than $D_{ze}$ and $Pi \mh D_{e}$. This is considerably different compared with the Harris current-sheet case \citep{pezzi2021dissipation} in which $D_{ze}$ is the dominant energy-transfer proxy. However, this behavior is consistent with turbulent simulations \citep{pezzi2021dissipation} and with observations of turbulent reconnection \citep{bandyopadhyay2021energy}.       


\section{Discussion}
\label{sec:discussion}

The type of magnetic reconnection that occurs from a turbulent cascade \citep{servidio2010statistics, loureiro2020nonlinear,fadanelli2021energy,agudelo2021three} presents a more complex geometry of the diffusion region compared to its laminar counterpart. Likewise, the geometry of the regions with enhanced energy transport and transfer is more complex. Moreover, in a 3D geometry, the particle motion along the out of plane direction allows energy transfer that 2D geometry precludes. For instance, the agyrotropic patterns in magnetic islands of 2D reconnection \citep{scudder2008illuminating} are located in the diffusion region outside the magnetic islands. Conversely, in our 3D case we observe agyrotropic patterns in the cross section of the flux-ropes, which we call magnetic islands.   

Since the plasma density is greater in the centers of the magnetic islands, these regions exhibit a greater plasma pressure compared to outside the islands. Patterns of agyrotropic plasma pressure are present not only within the magnetic islands but also in the regions between them (Figure \ref{fig:pepi_PI}).

The non-uniform guide magnetic field present in this reconnection event affects its geometry. Despite the 3D nature of this event, for the diffusion region in which {$|\vec{B}| \leq 0.4B_{0}$}, we  observe gyrotropic/agyrotropic patterns (section \ref{sec:Agirotropy}) similar to those observed in 2D laminar reconnection without guide field \citep{yin2001hybrid}. However, given the complex geometry of our event, we do not observe gyrotropic/agyrotropic patterns matching 2D reconnection in the part of the diffusion region below the x-points. Moreover, we do not observe a quadrupolar pattern of the in-plane component $\Pi_{pr,e}$    

\makeatletter\onecolumngrid@push\makeatother
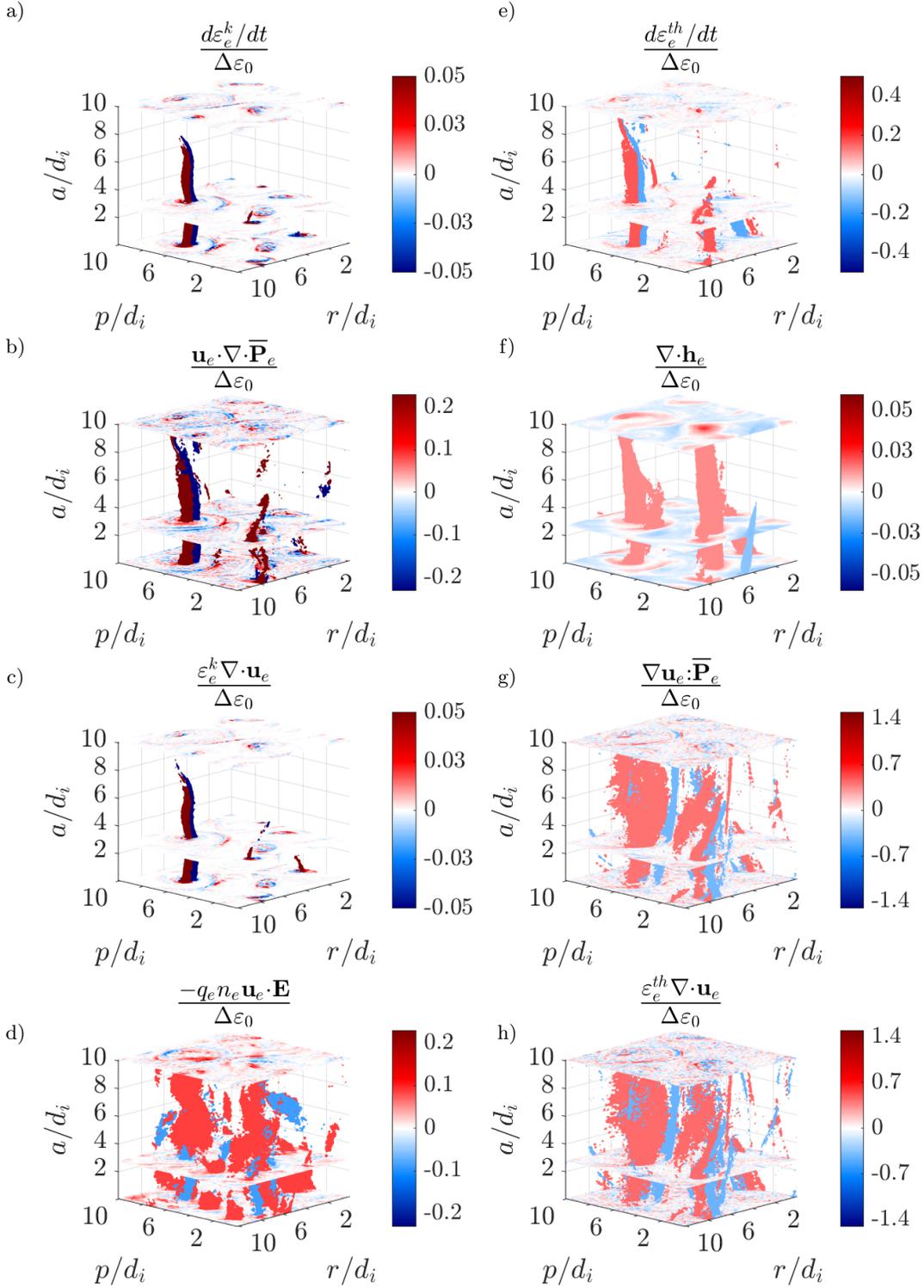
\begin{figure*}
\centering
\begin{tikzpicture}
    \draw (0, 0) node[inner sep=0] 
    {\includegraphics[width=0.87\linewidth]{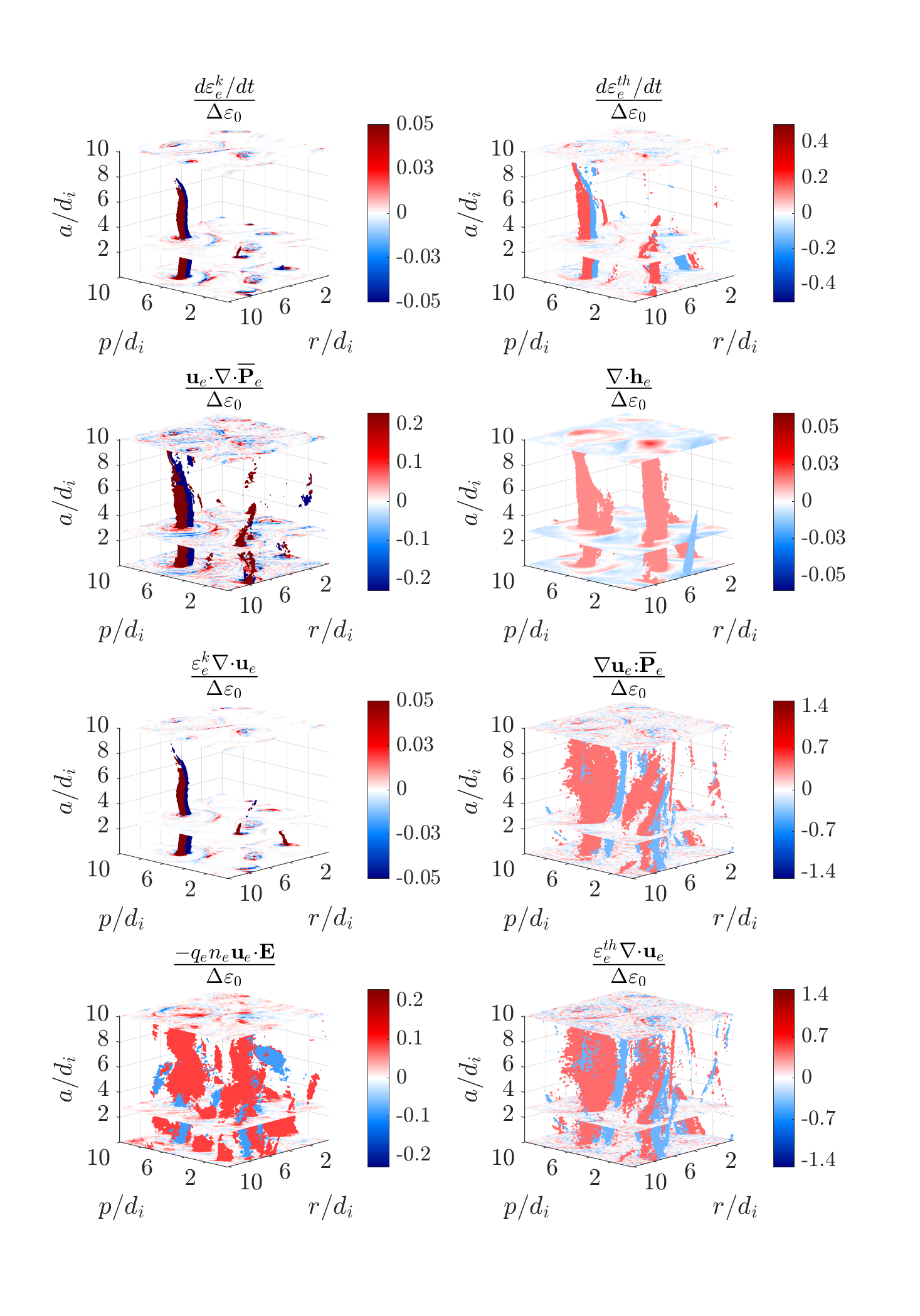}};
    \draw (-7.4, 10.1) node {a)};
    \draw (-7.4, 4.9) node {b)};
    \draw (-7.4, -0.2) node {c)};
    \draw (-7.4, -5.7) node {d)};
    \draw (0.2, 10.1) node {e)};
    \draw (0.2, 4.9) node {f)};
    \draw (0.2, -0.2) node {g)};    
    \draw (0.2, -5.7) node {h)};
\end{tikzpicture}
\vspace*{-1.3cm}
\caption{2D cuts in the $rp$-plane and isosurfaces of the power density terms at the simulation time $t=120 \omega^{-1}_{pi}$. Panels a) to e): kinetic power density terms for electrons. Panels f) to j): thermal power density terms for electrons. All quantities are normalised to $\Delta \varepsilon_{0}=\omega_{pi}m_{i}v_{A,i}^{2}$.} 
\label{fig:power3d}
\end{figure*}

\begin{figure*}
\centering
\begin{tikzpicture}
    \draw (0, 0) node[inner sep=0] 
    {\includegraphics[width=0.8\linewidth]{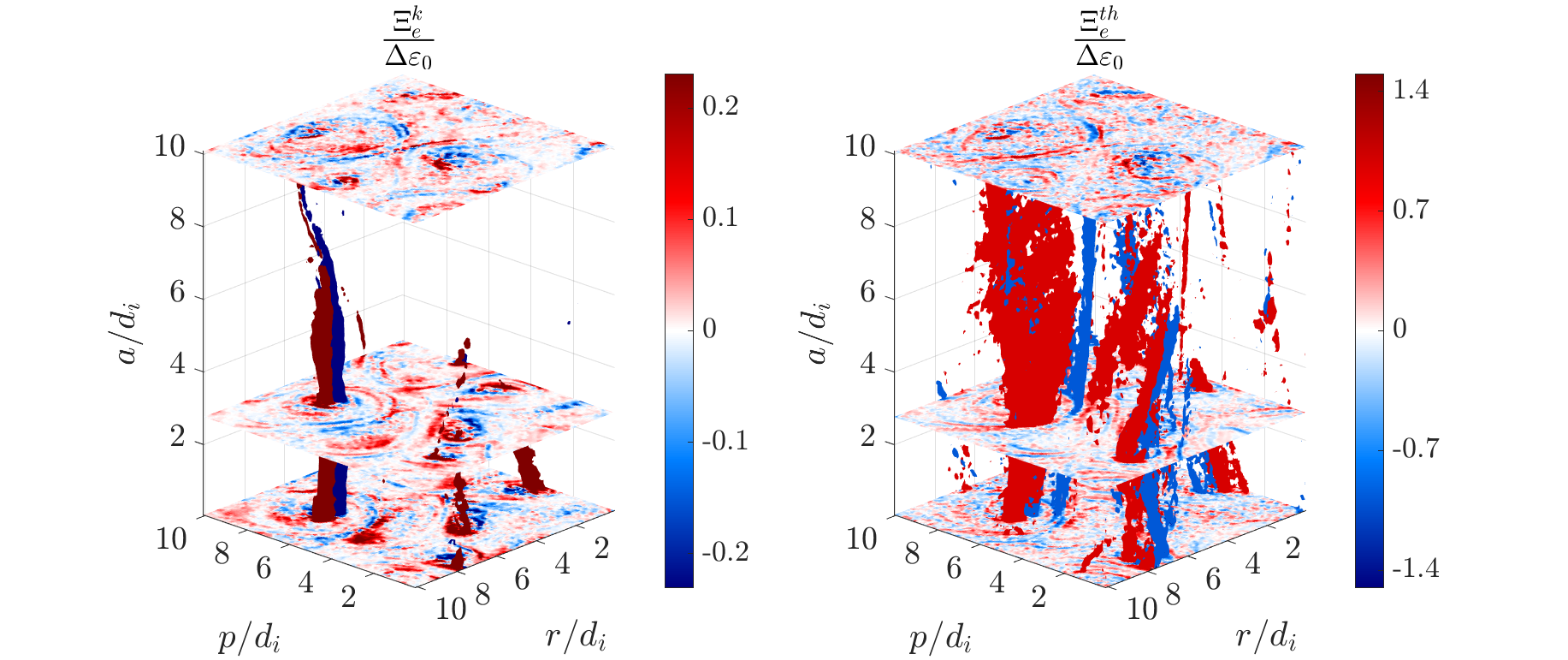}};
    \draw (-7.8, 3.5) node {a)};
    \draw (0.6, 3.5) node {b)};
\end{tikzpicture}
\caption{2D cuts in the $rp$-plane and isosurfaces at the simulation time $t=120 \omega^{-1}_{pi}$. Panel a): kinetic power density dissipation $\Xi^{k}_{e}$. b): thermal power density dissipation $\Xi^{th}_{e}$. All quantities are normalised to $\Delta \varepsilon_{0}=\omega_{pi}m_{i}v_{A,i}^{2}$.}
\label{fig:xikinther_e}
\end{figure*}


\begin{figure*}
\centering
{a)}
{\includegraphics[width=0.76\linewidth]{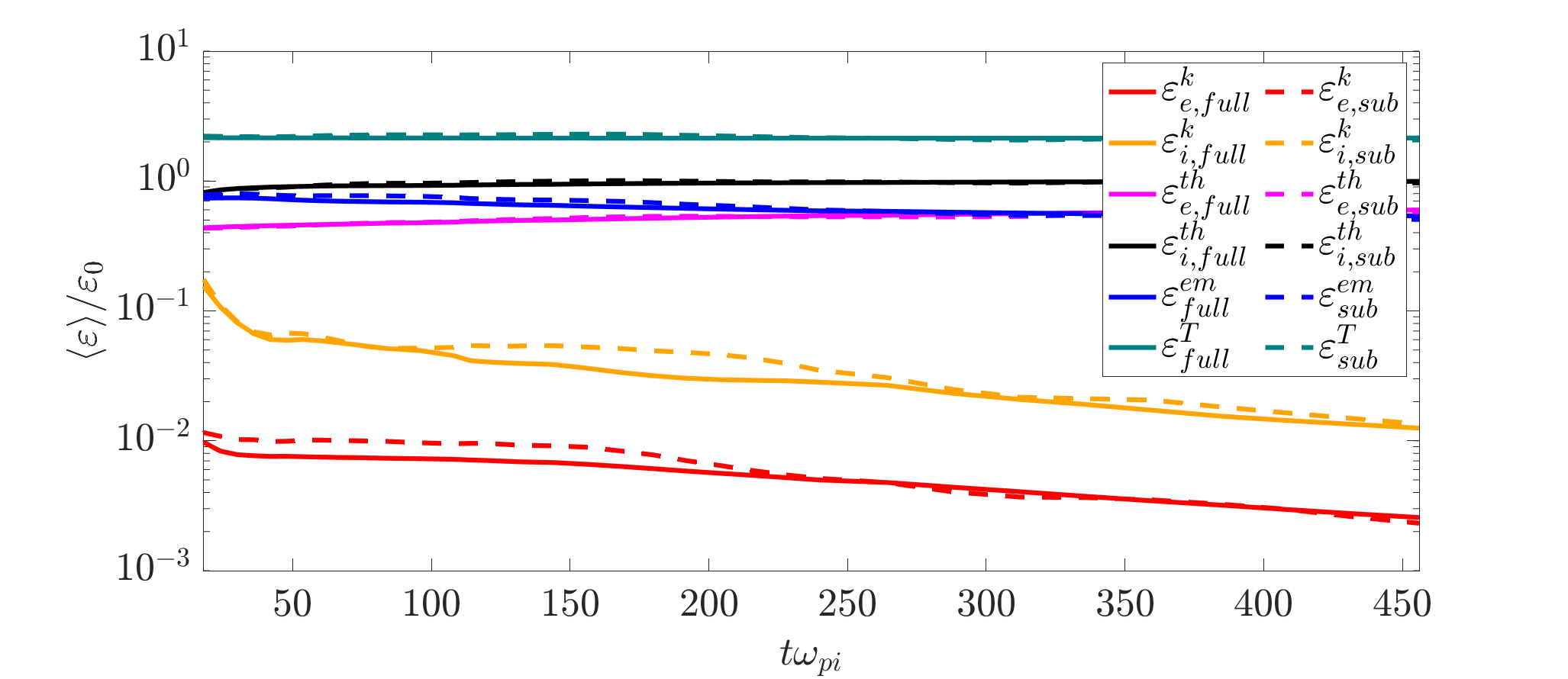}}\\
{b)}
{\includegraphics[width=0.76\linewidth]{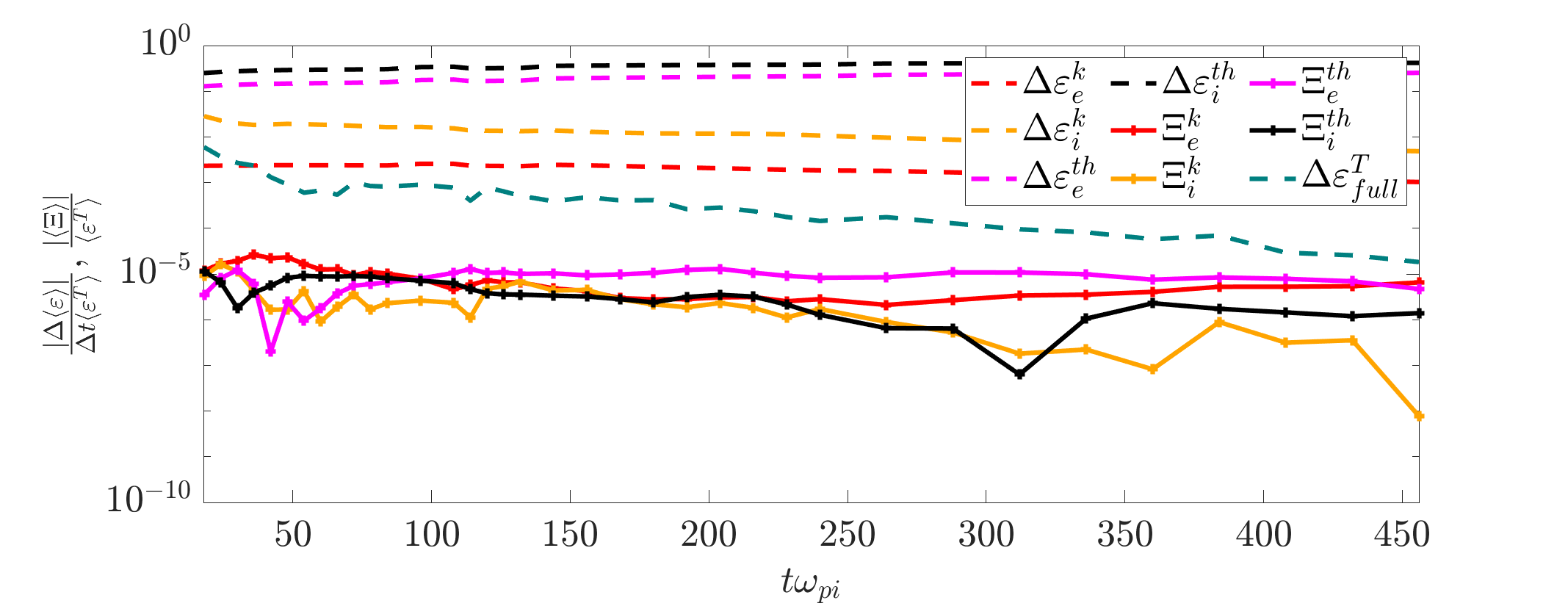}}\\
\caption{a) Time evolution of the energy densities averaged over the full simulation domain (solid curves) and over the sub-domain (dashed curves). The total energy density is $\varepsilon^{T}=\varepsilon^{k}_{e}+\varepsilon^{k}_{i}+\varepsilon^{th}_{e}+\varepsilon^{th}_{i}+\varepsilon^{em}$. b) Time evolution of the absolute values of the energy-density rates $\Delta \langle \varepsilon \rangle / \Delta t$ (dashed curves) and the dissipative power densities $\Xi$ (solid-dotted curves),  averaged over the full simulation domain and normalized to $\varepsilon^{T}_{full}$.} 
\label{fig:time_evolution}
\end{figure*}
\clearpage
\makeatletter\onecolumngrid@pop\makeatother

\begin{figure*}
\begin{tikzpicture}
\draw (0, 0) node[inner sep=0] 
{\includegraphics[width=1\linewidth]{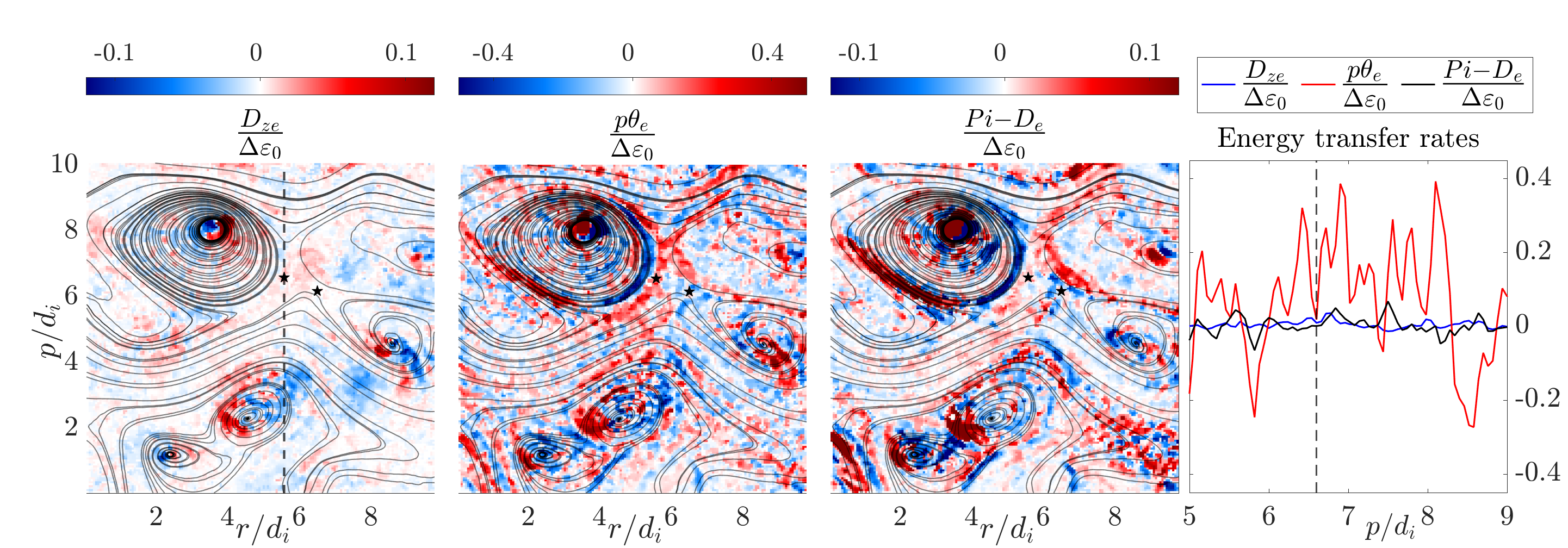}};
\draw (-7.8, 3.0) node {a)};
\draw (-3.4, 3.0) node {b)};
\draw (0.6, 3.0) node {c)};
\draw (5.0, 3.0) node {d)};
\end{tikzpicture}
\caption{Damping and heating proxies at the simulation time $t=120 \omega^{-1}_{pi}$. 2D cuts in the $rp$-plane of a) Zenitani parameter for electrons $D_{ze}$. b) Diagonal part of the strain-pressure interaction $p\theta_{e}$. c) Off-diagonal part of the strain-pressure interaction $Pi \mh D_{e}$. d) 1D cut of these terms as in Figure \ref{fig:power_ei_1D}.} 
\label{fig:D_pepi_PI_22}
\end{figure*}
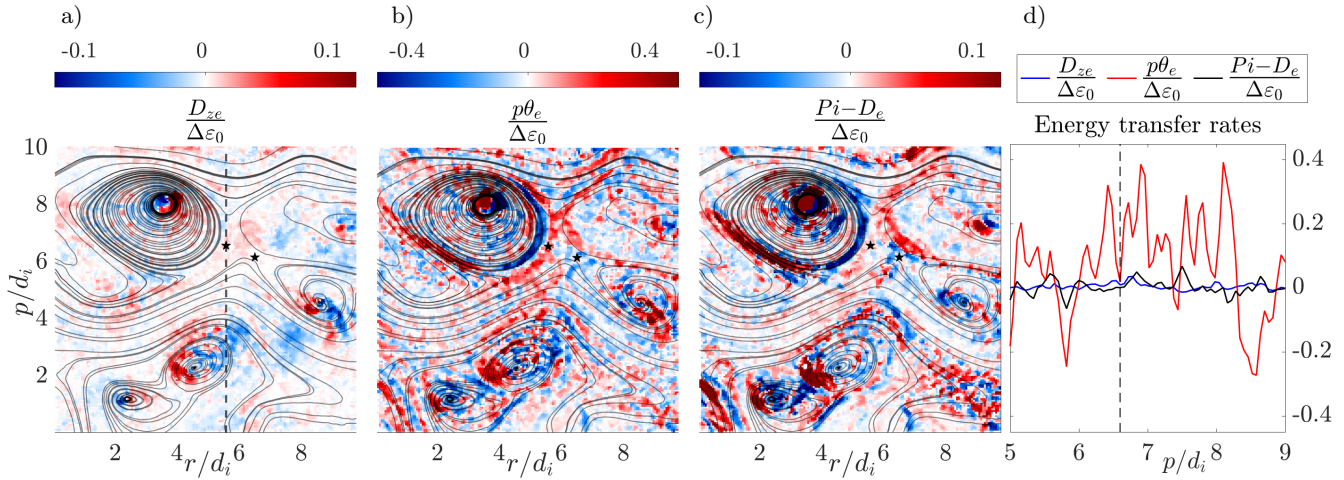

\noindent (Figure \ref{fig:pe_PI_12}d) within the diffusion region which is characteristic of agyrotropy in 2D reconnection without guide field \citep{yin2001hybrid}.

In the reconnection event that we analyze, although of turbulent nature, the out of plane electron motion is consistent with the 3D shape of electron diffusion regions observed in laboratory plasmas \citep{furno2005coalescence,yoo2013observation,yamada2014conversion}. 

In our event, ${d\varepsilon^{k}_{e}}/{dt}>0$ along the separatrices and ${d\varepsilon^{k}_{e}}/{dt}<0$ in the outer part of the reconnecting magnetic island (Figure \ref{fig:power_e}a). This corresponds to the acceleration of electrons along the separatrices (Figure \ref{fig:power_e}c) and the presence of a stagnation region. The shear between the flux ropes increases the electron thermal energy and pressure and the bulk kinetic energy reduces at the stagnation point. 

At the locations of the separatrices $\vec{u}_{e}\cdot \nabla \cdot \overline{\vec{P}}_{e}>0$ (Figure \ref{fig:power_e}b). This suggests electron streams that increase the electron pressure. Conversely, $\vec{u}_{e}\cdot \nabla \cdot \overline{\vec{P}}_{e}<0$ in the region between the x-points. This suggests electron streams that reduce the electron pressure and push the plasma within the diffusion region to a local thermal equilibrium. While reconnection is occurring, the high pressure electrons are allowed to fill in the diffusion region.

Within the diffusion region, the electric field increases the electron kinetic energy density and the work done by the electric field on the electrons $-q_{e}n_{e}(\vec{u}_{e}\cdot\vec{E})$ partially balances with the advection of the electron pressure. This is consistent with previous studies \citep{fadanelli2021energy}.   

The irreversible electron energy-density change $\Xi^{k}_{e}$ (Figure \ref{fig:power_e}e) is non-zero everywhere in the vicinity of the reconnecting structures. The quantity $\Xi^{k}_{e}$ displays structures with positive and negative values within the magnetic islands suggesting that collisional processes accelerate and decelerate electron bulk flows within the magnetic islands. Conversely, in the diffusion region, the positive value of $\Xi^{k}_{e}$ indicates that electrons are irreversibly accelerated. 

Unlike previous studies of turbulent reconnection \citep{fadanelli2021energy}, we estimate the electron thermal energy transfer associated with each term of Eq. (\ref{eqn:secondmomener_text}). Compared to the case of the kinetic power density, the thermal power density terms present stronger fluctuations. This is evident when comparing ${d\varepsilon^{k}_{e}}/{dt}$ (Figure \ref{fig:power_e}a) and ${d\varepsilon^{th}_{e}}/{dt}$ (Figure \ref{fig:power_e}f) as well as comparing $\varepsilon^{k}_{e}\nabla\cdot\vec{u}_{e}$ (Figure \ref{fig:power_e}c) and $\varepsilon^{th}_{e}\nabla\cdot\vec{u}_{e}$ ( Figure \ref{fig:power_e}i). This difference suggests that the electron bulk flows more efficiently transport thermal energy density than bulk kinetic energy density.

The power density terms associated with the compression/expansion of the flow $\nabla \vec{u}_{e}:\overline{\vec{P}}_{s}$ and  $\varepsilon^{th}_{e}\nabla\cdot\vec{u}_{e}$ exhibit a strong coherence with the electron motion along the reconnection separatrices. The electron streams gain thermal energy (i.e., heating) associated with the reconnection. This is consistent with simulations of fast collisionless reconnection at low $\beta$ \citep{loureiro2013fast} and observations of magnetospheric reconnection \citep{chasapis2017electron,holmes2019structure}.    

The most important contribution to $\nabla \vec{u}_{e}:\overline{\vec{P}}_{s}$ comes from the isotropic part of the strain pressure term. Correspondingly, $\varepsilon^{th}_{e}\nabla\cdot\vec{u}_{e}$ presents patterns similar to $\nabla \vec{u}_{e}:\overline{\vec{P}}_{s}$. Moreover, the contribution of the off-diagonal elements in $\nabla \vec{u}_{e}$ and $\overline{\vec{P}}_{e}$ to the thermal energy transport is less than the isotropic contribution, which is consistent with previous studies of turbulent reconnection \citep{fadanelli2021energy,bandyopadhyay2021energy}. The terms associated with electron compressibility ($\varepsilon^{th}_{e}\nabla\cdot\vec{u}_{e}$ and $\nabla \vec{u}_{e}:\overline{\vec{P}}_{s}$) are typically greater than the heat flux contribution ($\nabla \cdot \vec{h}_{s}$) suggesting that, for collisionless reconnection, compressible thermal energy-density transport is important for electrons.    
{On average, within the sub-domain, the electrons gain kinetic energy at the expense of the electric field (Figure~\ref{fig:power3d}d). The electrons both lose and gain thermal energy (Figures~\ref{fig:power3d}g and \ref{fig:power3d}h) predominantly along thin sheet-like structures.}

Similar to the irreversible kinetic energy-density transfer $\Xi^{k}_{e}$, the irreversible thermal energy transfer $\Xi^{th}_{e}$ is non-zero within the reconnecting structures as well as within the diffusion region. Moreover, electrons irreversibly gain thermal energy density at the location of the separatrices and within the diffusion region. 

{The irreversible kinetic energy-density transfer is mainly confined to the flux-ropes in our simulation (Figure~\ref{fig:xikinther_e}a). Conversely, the irreversible thermal energy-density transfer (Figure~\ref{fig:xikinther_e}b) occurs in thin sheet-like structures that extend for over $5d_{i}$.}

Although $\langle \Xi^{k}_{e} \rangle$ is negligible compared to $\Delta \varepsilon^{k}_{e}$ (Figure~\ref{fig:time_evolution}b), the fact that $\Xi^{k}_{e}$ is comparable to $q_{e}n_{e}\vec{E}\cdot \vec{u}_{e}$ and $\vec{u}_{e} \cdot \nabla \cdot \overline{\vec{P}}_{e}$ (Figure~\ref{fig:power_ei_1D}a) implies that $\Xi^{k}_{e}$ must be considered in the local kinetic energy transfer of electrons as it includes important information about the oscillating energy associated with instantaneous field--particle correlations \citep{klein2016measuring,howes2017diagnosing,klein2017diagnosing}. Only meaningful averages of the non-linear correlations between the fluctuating electric field and the fluctuating perturbation of the distribution function define the secular transfer of energy from the fields to the particles. Therefore, we propose that an energy-balance analysis based on the energy-density expressions derived from the collisionless Vlasov equation is not entirely accurate for kinetic simulations. {Because numerical effects in kinetic simulations act as an effective collision operator, the energy balance equations derived from Vlasov without provision for the terms on the right-hand side of Eq. \ref{eqn:boltz} are not exactly satisfied.}

Comparing our results with damping ($D_{z,e}$) and heating ($p\theta_{e}$ and $Pi \mh D_{e}$) proxies \citep{pezzi2021dissipation}, we observe that fluctuations of $p\theta_{e}$ inside the diffusion region (Figure \ref{fig:D_pepi_PI_22}d) are typically greater than fluctuations of $D_{z,e}$ and $Pi \mh D_{e}$. {Integrating over the sub-domain (not shown here), we find that $p\theta_{e}/\Delta\varepsilon_{0}|_{V} > |PiD_{e}|/\Delta\varepsilon_{0}|_{V}$. This suggests that, within the sub-domain, the electron  heating is mostly due to compressive effects.} This is consistent with results from turbulent simulations \citep{pezzi2021dissipation} and observations of turbulent reconnection \citep{bandyopadhyay2021energy}, but not with results from simulations of laminar reconnection. 

The proxies $p\theta_{e}$  and $Pi \mh D_{e}$ share the same signs at most locations in our simulation domain. However, in the diffusion region near the null region, the opposite sign of $Pi \mh D_{e}$ and $p\theta_{e}$ suggests that agyrotropic heating mechanisms can emerge to compensate for any reduction or increase in the thermal energy density due to isotropic heating mechanisms. {Moreover, integrating over the sub-domain and over time, we find that $p\theta_{e}|_{V,t}=0.0137$ and $Pi \mh D_{e}|_{V,t}=-0.0190$. This suggests that $p\theta$ is greater than $Pi\mh D$ within the diffusion region {at the particular time selected but not thought out the whole simulation}, due to a local effect.}

The positive values of $D_{z,e}$ and the negative value of $p\theta_{e}$ and $Pi \mh D_{e}$ in the region between the x-points suggest that electrons gain kinetic energy density from the fields while losing thermal energy density. {Between the x-points, the electric field accelerates electrons (Figure~\ref{fig:power_e}d). The increase in the electrons' kinetic energy density may be due to Landau damping \citep{landau1946oscillations,howes2006astrophysical,li2016energy}. Conversely, the magnetic pressure (not shown here) increases near the region between the x-points. The total pressure balance requires a depletion of $p_{e}$ and $p_{i}$ (as confirmed by Figures~\ref{fig:pepi_PI}a and \ref{fig:pepi_PI}e) in the diffusion region. Plasma pressure depletion have been suggested to be responsible for the onset of fast reconnection in collisionless plasmas \citep{liu2022first}. Thus the expansion and the consequential cooling-off of the electrons reduces their thermal energy. }


\section{Conclusions}
\label{sec:conclusions}
We derive a framework to quantify the collision-like effects that lead to irreversible energy transfer and thus dissipation in PIC plasmas. We identify and locate magnetic reconnection as a key mechanism for heating, damping, and dissipation in plasma turbulence in low-collisionality systems like the solar wind.  

Previously, the transfer and transport of energy in plasmas with low collisionality has been studied separately in simulations of reconnection \citep{hesse1998electron,hesse2001collisionless,zenitani2011new,munoz2017turbulent,pucci2018energy,pezzi2019energy,pezzi2021dissipation} and turbulence \citep{wan2012intermittent,yang2017energy,li2019collisionless,pezzi2021dissipation}. The transfer and transport in magnetic reconnection that forms from a turbulent cascade have been limited to 2D geometries \citep{parashar2009kinetic,fadanelli2021energy} and observations \citep{chasapis2018energy,bandyopadhyay2020statistics}, while the 3D case has received little attention. We study, for the first time the energy transport associated with 3D magnetic reconnection that occurs as a consequence of a turbulent cascade to a high level of detail and including all power density terms resulting from the full Boltzmann equation. We extend the analysis of similar studies \citep{fadanelli2021energy} by exploring the transfer and transport of thermal energy for electrons.

The energy transfer and transport in collisionesless plasmas is believed to be governed by non-thermal and kinetic mechanisms such as resonant \citep{marsch2003ion, kasper2008hot} and non resonant heating processes \citep{chandran2010perpendicular,chandran2013stochastic}. However, the irreversible energy transport is ultimately associated with collisional effects \cite{schekochihin2009astrophysical}. 

The agyrotropy signatures present in the reconnection diffusion region as well as in the reconnecting magnetic structures allows for agyrotropic energy transfer mechanisms such agyrotropic-driven instabilities to take place not only near the electron diffusion region \citep{ricci2004influence,roytershteyn2012influence,graham2017instability} but also within the reconnecting magnetic structures. {These signatures are three-dimensional as they extend in the $a$-direction for over $5d_{i}$}. A study of the instabilities that occur during a 3D turbulent reconnection event would be worthwhile to enhanced our understanding of the collisionless energy dissipation. 

We show that the contribution to the energy-density transfer from collision is not negligible. To determine the exact source of this contribution, future work must use large number of particles while keeping the 3D geometry. In addition, the inclusion of a controllable collision operator would allow for a detailed study of collisions in 3D reconnection \citep{pezzi2017solar,donnel2019multi,boesl2020collisional,pezzi2021dissipation}.   

The general framework that we introduce is suitable for estimating the irreversible energy-density transfer of the particle species in the solar wind. For instance, Eqs. (\ref{eqn:firstmomener_text}) and (\ref{eqn:secondmomener_text}) can be applied to spacecraft data to study the radial evolution of energy as a function of heliospheric distance in the solar wind. This work would be of interest both for the energetics of solar-wind electrons \citep{scime1994regulation,innocenti2020collisionless} and the solar wind prontons \citep{matteini2007evolution,hellinger2011heating,adhikari2020turbulence}

\acknowledgments


J.A.A.R~is supported by the European Space Agency's Networking/Partnering Initiative (NPI) program under contract 4000127929/19/NL/MH/mg and the Colombian program Pasaporte a la Ciencia, Foco Sociedad - Reto 3 under grant 3933061. D.V.~is supported by STFC Ernest Rutherford Fellowship  ST/P003826/1. D.V., G.N. and C.J.O.~are supported by STFC Consolidated Grants ST/S000240/1 and ST/W001004/1. R.T.W.~is supported by STFC Consolidated Grant ST/V006320/1. K.G.~is supported by NSF grant AGS-1460190. This work was performed using the DiRAC Data Intensive service at Leicester, operated by the University of Leicester IT Services, which forms part of the STFC DiRAC HPC Facility (www.dirac.ac.uk). The equipment was funded by BEIS capital funding via STFC Capital Grants ST/K000373/1 and ST/R002363/1 and STFC DiRAC Operations Grant ST/R001014/1. DiRAC is part of the National e-Infrastructure. This work was discussed at the ``Joint Electron Project'' at MSSL.

\appendix


\section{Derivation of the equations for the energy densities}
\label{app:Energy_equa}

To derive the $n$th moment of the Boltzmann equation (\ref{eqn:boltz}), we take the dyadic product of Eq. (\ref{eqn:boltz}) with $\vec{v}^{n}$ on the left and integrate the equation over the entire velocity space. The zeroth moment leads to  

\begin{equation}
 \frac{\partial n_{s} }{\partial t} + \nabla \cdot (n_{s}\vec{u}_{s}) = \Xi_{s}^{0}.
 \label{eqn:zero_moment}
\end{equation}

\noindent The collision operator has the property $\Xi_{s}^{0} = 0$ as it must conserve the number of particles. In this case, Eq. (\ref{eqn:zero_moment}) is the continuity equation. The first moment leads to 

\begin{equation}
    \frac{\partial (n_{s}\vec{u}_{s})}{\partial t} + \frac{1}{m_{s}} \nabla \cdot \tensnd{P}_{s} - \frac{q_{s}}{m_{s}}n_{s}(\vec{E} + \vec{u}_{s}\times\vec{B})  = \vec{\Xi}_{s}^{1}, 
    \label{eqn:frist-mom1}
\end{equation}

\noindent where 

\begin{equation}
\tensnd{P}_{s} \equiv m_{s}\int f_{s}\vec{v}\vec{v} d^{3}v.
\label{eqn:pressure_tensor2}
\end{equation}

\noindent We separate the second moment in Eq. (\ref{eqn:pressure_tensor2}) according to $ \nabla \cdot \tensnd{P} =  \nabla \cdot \tensndd{P} + \nabla \cdot (nm\vec{u}\vec{u})$, where $\tensndd{P}$ is defined in Eq. (\ref{eqn:pressure_tensor}). Invoking Eq. (\ref{eqn:zero_moment}), Eq. (\ref{eqn:frist-mom1}) takes the form

\begin{equation}
    \frac{d (n_{s}m_{s}\vec{u}_{s})}{d t}  =  - \nabla \cdot \tensndd{P}_{s} - (\nabla \cdot \vec{u}_{s})n_{s} m_{s}\vec{u}_{s} + q_{s}n_{s}(\vec{E} + \vec{u}_{s}\times\vec{B}) + m_{s}\vec{\Xi}_{s}^{1}.
    \label{eqn:firstmom}
\end{equation}

\noindent This equation describes the total change in time of the bulk momentum density for each species. 

The second moment of Eq. (\ref{eqn:boltz}) yields

\begin{eqnarray}
    \frac{\partial \tensndd{P}_{s}}{\partial t} + \nabla \cdot \left[({Q}_{ijk,s} + u_{i,s}{P}_{ij,s} + {P}_{ij,s}u_{k,s} + u_{j,s}{P}_{ik,s})\vec{\hat{e}^{i}}\otimes\vec{\hat{e}^{j}}\otimes\vec{\hat{e}^{k}}\right] - \frac{q_{s}}{m_{s}}\left( \tensndd{P}_{s}\times\vec{B} - \vec{B}\times \tensndd{P}_{s} \right) \nonumber = \\
    -\nabla \cdot\left(n_{s}m_{s}\vec{u}_{s}\vec{u}_{s}\vec{u}_{s}\right) - \frac{\partial (n_{s}m_{s}\vec{u}_{s}\vec{u}_{s})}{\partial t} + q_{s}n_{s}\left[ \vec{E}\vec{u}_{s} + \vec{u}_{s}\vec{E} + \frac{1}{m_{s}} \left( (\vec{u}_{s}\vec{u}_{s})\times\vec{B} - \vec{B}\times (\vec{u}_{s}\vec{u}_{s})\right) \right] + m_{s}\tensndd{\Xi}_{s}^{2},
    \label{eqn:secondmom2}
\end{eqnarray}

\noindent where $Q_{ijk,s}$ represent the elements of the heat-flux tensor 

\begin{equation}
\tensrdd{Q}_{s} \equiv m_{s}\int f_{s}(\vec{v}-\vec{u}_{s})(\vec{v}-\vec{u}_s)(\vec{v}-\vec{u}_{s}) d^{3}v.
\end{equation}

\noindent Eqs. (\ref{eqn:firstmom}) and (\ref{eqn:secondmom2}) are the exact first and second moments of Eq. (\ref{eqn:boltz}). 

We proceed to derive expressions for the energy densities $\varepsilon_{s}^{k}$ and $\varepsilon_{s}^{th}$. For this purpose, we take the scalar product of Eq. (\ref{eqn:firstmom}) with $\vec{u}_{s}$ which leads to Eq. (\ref{eqn:firstmomener_text}). To obtain an expression for the thermal energy $\varepsilon^{th}$, we take the trace of Eq. (\ref{eqn:secondmom2}). For the calculation of the trace of the cross product terms in Eq. (\ref{eqn:secondmom2}) we use an element-wise approach. If $\vec{A}$ is a vector and $\tensnd{M}$ is a second rank tensor, the cross product is defined as 
$\vec{A}\times\tensnd{M} = \epsilon_{lip}A_{i}M_{pq}\vec{\hat{e}^{l}}\otimes\vec{\hat{e}^{q}}$. It can be shown that $\tensnd{M} \times \vec{A} =-\left( \vec{A}\times \tensnd{M}^{T} \right)^{T}$, where $\tensnd{M}^{T}$ is the transposed of $\tensnd{M}$ and $Tr(\vec{A}\times\tensnd{M}) = \epsilon_{ijk}A_{i}M_{jk}$. Moreover, if $\tensnd{M}$ is a symmetric tensor, then $Tr(\vec{A}\times\tensnd{M}) = 0$. In addition, the trace of $\nabla \cdot \tensrdd{Q}$ corresponds to $2\nabla \cdot \vec{h}$. This procedure leads to 

\begin{eqnarray}
\frac{d \varepsilon_{s}^{th}}{d t} + \frac{d \varepsilon_{s}^{k}}{d t} + \nabla \cdot \vec{h}_{s} + \nabla\vec{u}_{s}:\tensndd{P}_{s} + (\nabla \cdot \vec{u}_{s})\varepsilon_{s}^{th} + \vec{u}_{s}\cdot(\nabla \cdot \tensndd{P}_{s}) = \nonumber \\
 - (\nabla \cdot \vec{u}_{s})\varepsilon_{s}^{k} + q_{s}n_{s}\vec{E}\cdot\vec{u}_{s} + \frac{1}{2}Tr\left( m_{s} \tensndd{\Xi}_{s}^{2}\right).
 \label{eqn:a20}
\end{eqnarray}

\noindent Combining Eqs. (\ref{eqn:firstmomener_text}) and (\ref{eqn:a20}), we obtain Eq. (\ref{eqn:secondmomener_text}).



\bibliography{Energy_transport_reconnection}{}
\bibliographystyle{aasjournal}

\end{document}